\documentstyle[aps,preprint]{revtex}

\def\[{\left\lbrack}
\def\]{\right\rbrack}

\def\({\left(}
\def\){\right)}

\newcommand{\be}{\begin{equation}}
\newcommand{\ee}{\end{equation}}
\newcommand{\ea}{\end{eqnarray}}
\newcommand{\ba}{\begin{eqnarray}}

\title{Embedding Second Class Systems via Symplectic Gauge-invariant Formalism}

\author{ A.C.R. Mendes\thanks{e-mail: albert@fisica.ufjf.br}}
\address{Centro Brasileiro  de Pesquisas F\'\i sicas,\\
Rua Xavier Sigaud 150, 22290-180, Rio de Janeiro, RJ, Brasil}

\author{J. Ananias Neto, W. Oliveira, C. Neves and D. C. Rodrigues\thanks{\noindent 
e-mail:jorge, wilson, cneves, davi@fisica.ufjf.br}} 
\address{Departamento de F\'{\i}sica,
ICE, Universidade Federal de Juiz de Fora,\\
36036-330, Juiz de Fora, MG, Brasil}

\begin{document}

\maketitle

\begin{abstract}
\noindent

In this paper we reformulate Abelian and non-Abelian noninvariant systems as gauge invariant theories using a new constraint conversion scheme, developed on the symplectic framework. This conversion method is not plagued by the ambiguity problem that torments the BFFT and iterative methods and also it seems more powerful since it does not require special modifications to handle with non-Abelian systems. 
\end{abstract}
   
\maketitle

\newpage

\setlength{\baselineskip} {20 pt}

\section{Introduction}

Gauge theories have played an important role in field theories since they are related with fundamental physical interactions on Nature.
In a more general sense, those theories have gauge symmetries defined by some relations called, in the Dirac's language, first class
constraints\cite{PD}. The quantization of these theories demand a special care because the presence of  gauge symmetries indicate that there are some superfluous degree of freedom, which must be eliminated (before or after) of the implementation of
a valid quantization process. The quantization of first class systems was formulated both in the Dirac\cite{PD} and path
integral\cite{FADDEEV} point of view. Later, the path integral analysis was extended by Batalin, Fradkin and Vilkovisky\cite{FRADKIN} in order to preserve
the BRST symmetry\cite{BRST}. 

On the other hand, the covariant quantization of second class systems is, in general, a difficult task because the Poisson brackets are
replaced by Dirac brackets. At the quantum level, the variables become operators and the Dirac brackets turn commutators. Due to this,
the canonical quantization process is tormented by serious problems, as ordering operator problems\cite{order} and anomalies\cite{RR} in
the context of nonlinear constrained systems and chiral gauge theories, respectively. In view of this, it seems that is more natural and
safe to work out the quantization of second class systems without invoking Dirac brackets. Actually, it was the strategy followed
by many authors over the last decades. The noninvariant system has been embeded in an extended phase space in order to change
the second class nature of constraints to first one. In this way, whole machinery\cite{BRST,BV} for quantizing
first class systems can be used as well. To implement this conception, Faddeev\cite{FS} suggests to enlarge the phase
space with the introduction of new variables to linearize the system, call Wess-Zumino(WZ) variables. This idea has been embraced by
many authors and some methods were proposed and some constraint conversion formalisms, based on the Dirac's method\cite{PD}, were
developed. Among them, the Batalin-Fradkin-Fradikina-Tyutin (BFFT)\cite{BT} and the iterative\cite{IJMP} methods were powerful enough to be successfully applied to a great
number of important physical systems. Although these techniques share the same conceptual basis \cite{FS} and follow the Dirac's
framework\cite{PD}, these constraint conversion methods were implemented following different directions. Historically, both BFFT and
the iterative methods were applied to deal with linear systems such as chiral gauge theories\cite{IJMP,many3} in order to eliminate
the gauge anomaly that hampers the quantization process. In spite of the great success achieved by these methods, they have an ambiguity
problem\cite{BN}. This problem naturally arise when the second class constraints are converted into first class ones with the introduction
of WZ variables. Due to this, the constraint conversion process may become a hard task, as shown in \cite{BN}.

The motivation of this paper is to introduce a new constraint conversion scheme that is not plagued by the ambiguity problem that torments
the BFFT and iterative methods and then to make a thorough investigation into the Hamiltonian formulation of the invariant model in order
to disclose the hidden symmetry of the model on the extended and original phase space.

We have organized this paper as follows. In section 2, we introduce the symplectic gauge-invariant formalism in order to settle the
notation and familiarize the reader with the fundamentals of the formalism. This formalism is developed on the symplectic
framework\cite{FJ,BC}, that is a modern way to handle with constrained systems. The basic object behind this
formalism is the symplectic matrix: if this matrix is singular, the model presents a symmetry, on the contrary, the Dirac brackets are
obtained. In view of this, we propose to render nonsingular symplectic matrix to a singular one. It will be done introducing arbitrary
functions dependent on the original and WZ variables into the first-order Lagrangian. To appreciate this point, a brief review of symplectic formalism will
be done and, after, general ideas of the symplectic gauge-invariant formalism will be presented. This formalism, on the contrary of BFFT
and iterative constraint conversion methods, does not require special modifications into the procedures to convert Abelian or non-Abelian set
of second class constraints into first class one. In section 3, we will make an application of the ideas discussed before in some Abelian models. We initiate with the Proca model in order to set up the general ideas discussed in Section 2. After that, we apply this
formalism in two important models to high energy physics. The first is the nonlinear sigma model (NLSM)\cite{NLSM}, which is an important
theoretical laboratory to learn the basic facts of life in asymptotically free field theories, as dynamical mass generation, confinement,
and topological excitations, which is expected in the realistic world of four-dimensional non-Abelian gauge theories. The second is the
bosonized chiral Schwinger model(CSM), which has attracted much attention over the last decade, mainly in the context of string
theories\cite{STRING}, and also due to the huge progress in understanding the physical meaning of anomalies in quantum field theories
achieved through the intensively study of this model. Through this section, we will first compute the Dirac brackets among the phase space
fields and, after, the gauge invariant version of the model will obtained. Later on, the gauge symmetry will be investigated from the
Dirac point of view. Section 4 is devoted to an application of the symplectic gauge-invariant formalism to the non-Abelian Proca model.
In this section, we will show that this gauge-invariant formalism does not require modifications to deal with non-Abelian models as demanded by the BFFT method. That is one of the great advantages of this method. Our concluding observations and final comments are given in Section 5.

\section{General formalism}

In this section, we briefly review the new gauge-invariant technique that changes the second class nature of a constrained system to first
one. This technique follows the Faddeev's suggestion\cite{FS} and is set up on a contemporary framework to handle constrained models,
namely, the symplectic formalism\cite{FJ,BC}. 

In order to systematize the symplectic gauge-invariant formalism, we consider a general noninvariant mechanical model whose dynamics is
governed by a Lagrangian ${\cal L}(a_i,\dot a_i,t)$(with $i=1,2,\dots,N$), where $a_i$ and $\dot a_i$ are the space and velocities
variables respectively. Notice that this model does not lead to lost generality or physical content. Following the symplectic method the
Lagrangian is written in its first-order form as
 
\begin{equation}
\label{2000}
{\cal L}^{(0)} = A^{(0)}_\alpha\dot\xi^{(0)}_\alpha - V^{(0)},
\end{equation}
where $\xi^{(0)}_\alpha(a_i,p_i)$(with $\alpha=1,2,\dots,2N$) are the symplectic variables, $A^{(0)}_\alpha$ are the one-form canonical
momenta, $(0)$ indicates that it is the zeroth-iterative Lagrangian and $V^{(0)}$ is the symplectic potential. After, the symplectic
tensor, defined as

\begin{eqnarray}
\label{2010}
f^{(0)}_{\alpha\beta} = {\partial A^{(0)}_\beta\over \partial \xi^{(0)}_\alpha}
-{\partial A^{(0)}_\alpha\over \partial \xi^{(0)}_\beta},
\end{eqnarray}
is computed. Since this symplectic matrix is singular, it has a zero-mode $(\nu^{(0)})$ that generates a new constraint when contracted
with the gradient of symplectic potential, namely,

\begin{equation}
\label{2020}
\Omega^{(0)} = \nu^{(0)}_\alpha\frac{\partial V^{(0)}}{\partial\xi^{(0)}_\alpha}.
\end{equation}
Through a Lagrange multiplier $\eta$, this constraint is introduced into the zeroth-iterative Lagrangian (\ref{2000}), generating the next
one,

\begin{eqnarray}
\label{2030}
{\cal L}^{(1)} &=& A^{(0)}_\alpha\dot\xi^{(0)}_\alpha + \dot\eta\Omega^{(0)}- V^{(0)},\nonumber\\
&=& A^{(1)}_\alpha\dot\xi^{(1)}_\alpha - V^{(1)},
\end{eqnarray}
where

\begin{eqnarray}
\label{2040}
V^{(1)}&=&V^{(0)}|_{\Omega^{(0)}= 0},\nonumber\\
\xi^{(1)}_\alpha &=& (\xi^{(0)}_\alpha,\eta),\\
A^{(1)}_\alpha &=& A^{(0)}_\alpha + \eta\frac{\partial\Omega^{(0)}}{\partial\xi^{(0)}_\alpha}.\nonumber
\end{eqnarray}
The first-iterative symplectic tensor is computed as

\begin{eqnarray}
\label{2050}
f^{(1)}_{\alpha\beta} = {\partial A^{(1)}_\beta\over \partial \xi^{(1)}_\alpha}
-{\partial A^{(1)}_\alpha\over \partial \xi^{(1)}_\beta}.
\end{eqnarray}
Since this tensor is nonsingular, the iterative process stops and the Dirac's brackets among the phase space variables are obtained from
the inverse matrix $(f^{(1)}_{\alpha\beta})^{-1}$. On the contrary, the tensor has a zero-mode and a new constraint arises, indicating
that the iterative process goes on.

After this brief review, the symplectic gauge-invariant formalism will be systematized. It starts with the introduction of two arbitrary
functions dependent on the original phase space and WZ variables, namely, $\Psi(a_i,p_i,\theta)$ and $G(a_i,p_i,\theta)$, into the first-order
Lagrangian as follows,

\be
\label{2060a}
{\tilde{\cal L}}^{(0)} = A^{(0)}_\alpha\dot\xi^{(0)}_\alpha + \Psi\dot\theta - {\tilde V}^{(0)},
\ee
with

\be
\label{2060b}
{\tilde V}^{(0)} = V^{(0)} + G(a_i,p_i,\theta),
\ee
where the arbitrary function, given by,

\begin{equation}
\label{2060}
G(a_i,p_i,\theta)=\sum_{n=0}^\infty{\cal G}^{(n)}(a_i,p_i,\theta),
\end{equation}
satisfies the following boundary condition

\begin{eqnarray}
\label{2070}
G(a_i,p_i,\theta=0) = {\cal G}^{(n=0)}(a_i,p_i,\theta=0)=0.
\end{eqnarray}
The symplectic variables were extended to also contain the WZ variable $\tilde\xi^{(1)}_{\tilde\alpha} = (\xi^{(0)}_\alpha,\theta)$
(with ${\tilde\alpha}=1,2,\dots,2N+1$) and the first-iterative symplectic potential becomes

\begin{equation}
\label{2075}
{\tilde V}_{(n)}^{(0)}(a_i,p_i,\theta) = V^{(0)}(a_i,p_i) + \sum_{n=0}^\infty{\cal G}^{(n)}(a_i,p_i,\theta).
\end{equation}
For $n=0$, we have

\begin{equation}
\label{2075a}
{\tilde V}_{(n=0)}^{(0)}(a_i,p_i,\theta) = V^{(0)}(a_i,p_i).
\end{equation}

The implementation of the symplectic gauge-invariant scheme follows two steps: the first one is addressed to compute $\Psi(a_i,p_i,\theta)$
while the second one is dedicated to the calculation of $G(a_i,p_i,\theta)$. To start the first step, we impose that the new symplectic
tensor (${\tilde f}^{(0)}$) is singular, then,

\begin{equation}
\label{2076}
\sum_{\tilde\alpha} \tilde\nu^{(0)}_{\tilde\alpha}{\tilde f}^{(0)\tilde\alpha\tilde\beta} = 0,
\end{equation}
where $\tilde\nu^{(0)}_{\tilde\alpha}$ is a zero-mode, reads as

\begin{equation}
\label{2076a}
\tilde\nu^{(0)}_{\tilde\alpha}=\pmatrix{\nu^{(0)}_\alpha & 1}.
\end{equation}
>From relation (\ref{2076}) some differential equations involving $\Psi(a_i,p_i,\theta)$ are obtained and, after a straightforward
computation, $\Psi(a_i,p_i,\theta)$ can be determined.

Afterward, the second step starts. In order to compute $G(a_i,p_i,\theta)$, we impose that no more constraints arise from the contraction
of the zero-mode $(\tilde\nu^{(0)}_{\tilde\alpha})$ with the gradient of potential ${\tilde V}_{(n)}^{(0)}(a_i,p_i,\theta)$. This condition
generates a general differential equation, reads as

\begin{eqnarray}
\label{2080}
\tilde\nu^{(0)}_{\tilde\alpha}\frac{\partial {\tilde V}_{(n)}^{(0)}(a_i,p_i,\theta)}{\partial{\tilde\xi}^{(0)}_{\tilde\alpha}}&=& 0,\nonumber\\
\nu^{(0)}_\alpha\frac{\partial V^{(0)}(a_i,p_i)}{\partial\xi^{(0)}_\alpha} + \sum_{n=0}^\infty\frac{\partial{\cal G}^{(n)}(a_i,p_i,\theta)}{\partial\theta}&=&0,
\end{eqnarray}
that allows us to compute all correction terms ${\cal G}^{(n)}(a_i,p_i,\theta)$ in order of $\theta$. For linear correction term, we have

\begin{equation}
\label{2090}
\nu^{(0)}_\alpha\frac{\partial V_{(n=0)}^{(0)}(a_i,p_i)}{\partial\xi^{(0)}_\alpha} + \frac{\partial{\cal
 G}^{(n=1)}(a_i,p_i,\theta)}{\partial\theta} = 0.
\end{equation}
For quadratic correction term, we get

\begin{equation}
\label{2095}
{\nu}^{(0)}_{\alpha}\frac{\partial V_{(n=1)}^{(0)}(a_i,p_i,\theta)}{\partial{\xi}^{(0)}_{\alpha}} + \frac{\partial{\cal G}^{(n=2)}(a_i,p_i,\theta)}{\partial\theta} = 0.
\end{equation}
>From these equations, a recursive equation for $n\geq 1$ is proposed as

\begin{equation}
\label{2100}
{\nu}^{(0)}_{\alpha}\frac{\partial V_{(n - 1)}^{(0)}(a_i,p_i,\theta)}{\partial{\xi}^{(0)}_{\alpha}} + \frac{\partial{\cal
 G}^{(n)}(a_i,p_i,\theta)}{\partial\theta} = 0,
\end{equation}
that allows us to compute each correction term in order of $\theta$. This iterative process is successively repeated up to the equation
(\ref{2080}) becomes identically null, consequently, the extra term $G(a_i,p_i,\theta)$ is obtained explicitly. Then, the gauge invariant
Hamiltonian, identified as being the symplectic potential, is obtained as

\begin{equation}
\label{2110}
{\tilde{\cal  H}}(a_i,p_i,\theta) = V^{(0)}_{(n)}(a_i,p_i,\theta) = V^{(0)}(a_i,p_i) + G(a_i,p_i,\theta),
\end{equation}
and the zero-mode ${\tilde\nu}^{(0)}_{\tilde\alpha}$ is identified as being the generator of an infinitesimal gauge transformation, given by
\begin{equation}
\label{2120}
\delta{\tilde\xi}_{\tilde\alpha} = \varepsilon{\tilde\nu}^{(0)}_{\tilde\alpha},
\end{equation}
where $\varepsilon$ is an infinitesimal parameter. 

In the next section, we will apply the symplectic gauge-invariant formalism in some second class constrained Hamiltonian systems.

\section{Gauge-invariant reformulation of second class systems}

\subsection{The Abelian Proca model}

It is pedagogical to implement the general ideas of the new gauge-invariant formalism set up in the previous section to particular models.
The analysis of these models elucidate and gives deeper insight into the general formalism. To this end, let us start with a simple Abelian
case which is the Proca model whose dynamics is controlled by the Lagrangian density, 

\begin{equation}
\label{Proca1}
{\cal L} = - \,\,\frac{1}{4}F_{\mu\nu}F^{\mu\nu} + \frac{1}{2}\,m^2\,A^{\mu}A_{\mu},
\end{equation}
where $m$ is the mass of the $A_{\mu}$ field, $g_{\mu\nu} = diag(+,-)$ and $F_{\mu\nu} = \partial_{\mu}A_{\nu} - \partial_{\nu}A_{\mu}$.
Observe that the mass term breaks the gauge invariance of the usual Maxwell's theory. Hence, the Lagrangian density above represents a
second class system.

To perform the symplectic formalism the Lagrangian density is reduced to its first-order form, reads as

\begin{equation}
\label{Proca3}
{\cal L}^{(0)} = \pi^{i}\dot{A_{i}} - V^{(0)},
\end{equation}
where the symplectic potential is

\be
\label{Proca3a}
V^{(0)} = \frac{1}{2}{\pi_{i}}^2 + \frac 14 F_{ij}^2 + \frac{1}{2}\,m^2\,{A_{i}}^2 - A_{0}(\partial_i\pi^{i} + \frac{1}{2}\,m^2\,A_{0}) ,
\ee
with $\pi_i=\dot A_i - \partial_iA_0$, where $\partial_i=\frac{\partial}{\partial x^i}$ and dot denote space and time derivatives,
respectively. The symplectic fields are $\xi_\alpha^{(0)}=(A_i,\pi_i,A_0)$ with the corresponding one-form canonical momenta given by

\begin{eqnarray}
\label{Proca4}
a_{A_i}^{(0)} &=& \pi_i, \nonumber \\
a_{\pi^i}^{(0)} = a_{A_0}^{(0)} &=& 0.
\end{eqnarray}
The zeroth-iterative symplectic matrix is

\begin{equation}
f^{(0)} = \left(
\begin{array}{ccc}
0           & -\delta_{ij} & 0 \\
\delta_{ji}&         0     & 0 \\
0           &         0     & 0
\end{array}
\right)\,\delta^{(3)}({\vec x} - {\vec y}),
\end{equation}
that is a singular matrix. It has a zero-mode that generates the following constraint,

\begin{equation}
\label{Proca5}
\Omega = \partial_i\pi^i + m^2A_0,
\end{equation}
identified as being the Gauss's law. Bringing back this constraint into the canonical part of the first-order Lagrangian ${\cal L}^{(0)}$
using a Lagrangian multiplier ($\beta$), the first-iterated Lagrangian, written in terms of $\xi_\alpha^{(1)} = (A_i,\pi_i,A_0, \beta)$ is
obtained as

\begin{equation}
\label{Proca6}
{\cal L}^{(1)} = \pi^{i}\dot{A_{i}} + \Omega\dot{\beta} - V^{(1)},
\end{equation}
with the following symplectic potential,

\be
\label{Proca6a}
V^{(1)} = \frac{1}{2}{\pi_{i}}^2 + \frac 14 F_{ij}^2 + \frac{1}{2}\,m^2\,\({A_{0}}^2 + {A_{i}}^2\) - A_0\Omega.
\ee
The first-iterated symplectic matrix, computed as

\begin{equation}
f^{(1)}=\left(
\begin{array}{cccc}
0           & -\delta_{ij}  &  0   &   0 \\
\delta_{ji} &         0     &   0    &    \partial^y_i \\
0         &         0     &   0    &     m^2   \\
0       &        -\partial^x_j    &  -m^2     &     0  
\end{array}
\right)\,\delta^{(3)}({\vec x} - {\vec y}),
\end{equation}
is a nonsingular matrix and, consequently, the Proca model is not a gauge invariant field theory. As settle by the symplectic formalism, the Dirac brackets among the phase space fields are acquired from the inverse of the symplectic matrix, namely,

\ba
\label{dirac01}
\lbrace A_i(\vec x),A_j(\vec y)\rbrace^* &=& 0,\nonumber\\
\lbrace A_i(\vec x),\pi_j(\vec y)\rbrace^* &=& \delta_{ij}\delta^{(3)}(\vec x - \vec y),\\
\lbrace \pi_i(\vec x),\pi_j(\vec y)\rbrace^* &=& 0,\nonumber\\
\ea
with the following Hamiltonian,

\ba
\label{hamil01}
{\cal H} = V^{(1)}|_{\Omega=0} &=& \frac 12 \pi_i^2 - \frac {1}{2 m^2}\pi_i\partial^i\partial_j\pi^j + \frac 14 F_{ij}^2 + \frac 12 m^2 A_i^2,\nonumber\\
&=& \frac 12 \pi_iM^i_j\pi^j + \frac 14 F_{ij}^2 + \frac 12 m^2 A_i^2,
\ea
where the phase space metric is

\be
\label{metric01}
M^i_j = g^i_j - \frac{\partial^i\partial_j}{m^2}.
\ee
It completes the noninvariant analysis.

At this point we are ready to carry out the symplectic gauge-invariant formulation of the Abelian Proca model in order to disclose the gauge symmetry present on the model. To this end, we will extend the symplectic gauge-invariant formalism\cite{ANO}, recently proposed by three of us in
order to unveil the gauge symmetry present on the Skyrme model. The basic concept behind the extended symplectic gauge-invariant formalism dwells on the extension of the original phase space with the introduction of two arbitrary functions, $\Psi$ and $G$, depending on the original phase space variables and the WZ variable $(\theta)$. The former is introduced into the kinetical sector and, the later, into the potential sector of the first-order Lagrangian. The process starts with the computation of $\Psi$ and finishes with the computation of $G$.

In order to reformulate the Proca model as a gauge invariant field theory, let us start with the first-order Lagrangian ${\cal L}^{(0)},$ given in Eq.(\ref{Proca3}), with the arbitrary terms, given by,

\begin{equation}
\label{Proca6b}
{\tilde{\cal L}}^{(0)} = \pi^{i}\dot{A_{i}} + \dot\theta\Psi - {\tilde V}^{(0)},
\end{equation}
with

\be
\label{Proca6c}
{\tilde V}^{(0)} =  \frac{1}{2}{\pi_{i}}^2 + \frac 14 F_{ij}^2  + \frac{1}{2}\,m^2\,{A_{i}}^2 - A_{0}(\partial_i\pi^{i} + \frac{1}{2}\,m^2\,A_{0})  + G,
\ee
where $\Psi\equiv\Psi(A_i,\pi_i,A_0,\theta)$ and $G\equiv G(A_i,\pi_i,A_0,\theta)$ are arbitrary functions to be determined. Now, the symplectic fields are ${\tilde\xi}^{(0)}_\alpha=(A_i,\pi_i,A_0,\theta)$ while the symplectic matrix is

\be
\label{matrix00}
f^{(0)} = \pmatrix{ 0 & - \delta_{ij} & 0 & \frac{\partial\Psi_y}{\partial A^x_i}\cr \delta_{ji} &  0 & 0 & \frac{\partial\Psi_y}{\partial \pi^x_i}\cr 0 & 0 & 0 & \frac{\partial\Psi_y}{\partial A^x_0}\cr - \frac{\partial\Psi_x}{\partial A^y_j} & - \frac{\partial\Psi_x}{\partial \pi^y_j} & - \frac{\partial\Psi_x}{\partial A^y_0} & f_{\theta_x\theta_y}}\delta^{(3)}(\vec x - \vec y),
\ee
with

\be
\label{matrix01}
f_{\theta_x\theta_y} = \frac{\partial \Psi_y}{\partial \theta_x} - \frac{\partial \Psi_x}{\partial \theta_y},
\ee
where $\theta_x \equiv \theta(x)$, $\theta_y \equiv \theta(y)$, $\Psi_x \equiv \Psi(x)$ and $\Psi_y \equiv \Psi(y)$.

In order to unveil the hidden $U(1)$ gauge symmetry in the Proca model, the symplectic matrix above must be singular, then, $\Psi\equiv(A_i,\pi_i,\theta)$. As established by the symplectic gauge-invariant formalism, the corresponding zero-mode $\nu^{(0)}(\vec x)$, identified as being the generator of the symmetry, satisfies the relation below,

\be
\label{matrix02}
\int \,\, d^3y \,\,\nu^{(0)}_\alpha(\vec x)\,\,f_{\alpha\beta}(\vec x - \vec y)= 0,
\ee 
producing a set of equations that allows to determine $\Psi$ explicitly. At this point, it is very important to notice that the extended symplectic gauge-invariant formalism opens up the possibility to extract the gauge symmetry of the physical model, because the zero-mode does not generate a new constraint, however, it determines the arbitrary function $\Psi$ and, consequently, awards the gauge invariant reformulation of the model. We consider to scrutinize the gauge symmetry related to the following zero-mode,

\be
\label{matrix03}
\bar\nu^{(0)} = \pmatrix { \partial_i & 0 & 0 & 1}.
\ee
Since this zero-mode and the symplectic matrix (\ref{matrix00}) must satisfy the gauge symmetry condition given in Eq.(\ref{matrix02}), a set of equations is obtained and, after an integration process, $\Psi$ is computed as

\be
\label{matrix04}
\Psi = - \partial_i\pi^i.
\ee
In view of this, the symplectic matrix becomes

\be
\label{matrix05}
f^{(0)} = \pmatrix{ 0 & - \delta_{ij} & 0 & 0\cr \delta_{ji} &  0 & 0 & - \partial_i^y \cr 0 & 0 & 0 & 0 \cr 0 & \partial_j^x & 0 & 0}\delta^{(3)}(\vec x -\vec y), 
\ee
which is singular by construction. Due to this, the first-order Lagrangian is 

\begin{equation}
\label{Proca6ba}
{\tilde{\cal L}}^{(0)} = \pi^{i}\dot{A_{i}} - \partial_i\pi^i\dot\theta - {\tilde V}^{(0)},
\end{equation}
with ${\tilde V}^{(0)}$ given in Eq.(\ref{Proca6c}).

Now, let us to start with the second step to reformulate the model as a gauge theory. The zero-mode $\bar\nu^{(0)}$ does not produce a
constraint when contracted with the gradient of symplectic potential, namely,

\be
\label{matrix06}
\nu^{(0)}_\alpha\frac{\partial {\tilde V}^{(0)}}{\partial {\tilde\xi}_\alpha} = 0,
\ee
oppositely, it produces a general equation that allows to compute the correction terms in $\theta$ enclosed into $G(A_i,\pi_i,A_0,\theta)$,
given in Eq.(\ref{2080}). To compute the correction term linear in $\theta$, namely, ${\cal G}^{(1)}$, we pick up the following terms from
the general relation (\ref{2080}), given by,

\be
\label{matrix07}
\int_x \,\[\partial^w_l\(m^2A^l(x)\delta^{(3)}(\vec x - \vec w) + \frac 12 F_{ij}(x)\frac{\partial F^{ij}(x)}{\partial A^l(w)}\) +
\frac{\partial{\cal G}^{(1)}(x)}{\partial\theta(w)}\] = 0.
\ee
After a straightforward calculation, the correction term linear in $\theta$ is obtained as

\be
\label{matrix08}
{\cal G}^{(1)}  = - m^2\partial^iA_i \theta.
\ee
Bringing back this result into the symplectic potential (\ref{Proca6c}), we get

\be
\label{matrix08a}
{\tilde V}^{(0)} =  \frac{1}{2}{\pi_{i}}^2 + \frac 14 F_{ij}^2 + \frac{1}{2}\,m^2\,{A_{i}}^2 - A_{0}(\partial_i\pi^{i} +
\frac{1}{2}\,m^2\,A_{0}) - m^2\partial^iA_i\theta .
\ee
However, the invariant formulation of the Proca model was not yet obtained because the contraction of the zero-mode (\ref{matrix03}) with
the symplectic potential above does not generate a null value. Due to this, higher order correction terms in $\theta$ must be computed. For
quadratic term, we have,

\be
\label{matrix09}
\int_x\,\,\[ \partial^w_l\(- m^2\theta(x)\partial_x^l\delta^{(3)}(\vec x - \vec w)\)  +
\frac{\partial{\cal G}^{(2)}(x)}{\partial\theta(w)}\]= 0,
\ee
that after a direct calculation, we get

\be
\label{matrix10}
{\cal G}^{(2)}  = + \frac 12 m^2\,\, \,(\partial_i\theta)^2.
\ee
Then, the first-order Lagrangian becomes,

\be
\label{matrix10a}
{\tilde{\cal L}} = \pi^i\dot A_i + \dot\theta\Psi - {\tilde V}^{(0)},
\ee
where the symplectic potential is

\ba
\label{matrix11}
{\tilde V}^{(0)} &=& \frac{1}{2}{\pi_{i}}^2 + \frac 14 F_{ij}^2 + \frac{1}{2}\,m^2\,{A_{i}}^2 - A_{0}(\partial_i\pi^{i} +
\frac{1}{2}\,m^2\,A_{0}) - m^2\partial^iA_i\theta + \frac 12 m^2\,\, \,(\partial_i\theta)^2.
\ea

Since the contraction of the zero-mode $({\bar\nu}^{(0)})$ with the symplectic potential above does not produce a new constraint,
a hidden symmetry is revealed.
 
To complete the gauge invariant reformulation of the Abelian Proca model, the infinitesimal gauge transformation will be also computed.
In agreement with the symplectic method, the zero-mode $\bar \nu^{(0)}$ is the generator of the infinitesimal gauge transformation
$(\delta{\cal O}=\varepsilon\bar\nu^{(0)})$, given by,

\begin{eqnarray}
\label{Proca16}
\delta A_i &=& - \partial_i\varepsilon,\nonumber\\
\delta \pi_i &=& 0,\\
\delta A_0 &=& 0,\nonumber\\
\delta\theta &=& \varepsilon,\nonumber
\end{eqnarray}
where $\varepsilon$ is an infinitesimal time-dependent parameter. Indeed, for the above transformations the invariant Hamiltonian,
identified as being the symplectic potential ${\tilde V}^{(0)}$, changes as

\be
\label{matrix12}
\delta{\cal H} = 0.
\ee

At the present moment we are involved to investigate the result from the Dirac point of view. The chains of primary constraints computed
from the Lagrangian (\ref{matrix10a}) are

\ba
\label{matrix13}
\phi_1 &=& \pi_0,\nonumber\\
\chi_1 &=& \partial_i\pi^i + \pi_\theta .
\ea
Next, these constraints are introduced into the invariant Hamiltonian (\ref{matrix11}) through Lagrange multipliers and, then, it is
rewritten as

\be
\label{matrix14}
{\tilde V}^{(0)}_{primary} = {\tilde V}^{(0)} + \lambda_1\phi_1 + \gamma_1\chi_1.
\ee
The temporal stability condition for the primary constraint $\phi_1$ requires secondary constraint, reads as

\be
\label{matrix15}
\phi_2 = \partial_i\pi^i + m^2A_0,
\ee
and no more constraint appears from the time evolution of $\phi_2$. Now, the total Hamiltonian is written as

\be
\label{matrix16}
{\tilde V}^{(0)}_{total} = {\tilde V}^{(0)} + \lambda_1\phi_1 + \lambda_2\phi_2 + \gamma_1\chi_1.
\ee
Since the time evolution of $\phi_1$ just allows to obtain the Lagrange multiplier $\lambda_2$, and the constraint $\chi_1$ has no time evolution $\dot\chi_1=0$, no more constraints arise. Then, the gauge invariant model has three constraints $(\phi_1,\phi_2,\chi_1)$. The nonvanishing Poisson brackets among these constraints are,

\be
\lbrace \phi_1(x),\phi_2(y)\rbrace = - m^2\delta^{(3)}(\vec x - \vec y).
\ee
Then, the Dirac matrix, given by

\be
\label{matrix17}
C = \pmatrix{0 & -1  & 0\cr 1 & 0 & 0\cr 0 & 0 & 0}m^2\delta(\vec x - \vec y),
\ee
is singular, indicating that the model has, indeed, a gauge symmetry. However, it also has some nonvanishing Poisson brackets among the constraints, suggesting that the model has both first and second class constraints. It is easy to check that $\chi_1$ is a first class constraint and $\phi_1$ and $\phi_2$ are second class constraints. In accordance with the Dirac method, the set of second class constraint must be taken equal to zero in a strong way, generating then the primary Dirac brackets among the phase space fields, given by

\ba
\label{matrix19}
\lbrace A_i(\vec x), \pi_j(\vec y)\rbrace^* &=& \delta_{ij}\delta^{(3)}(\vec x - \vec y),\nonumber\\
\lbrace \theta(\vec x), \pi_\theta(\vec y)\rbrace^* &=& \delta^{(3)}(\vec x - \vec y).
\ea
The gauge invariant version of the Abelian Proca model is then governed by an invariant Hamiltonian, reads as

\ba
\label{matrix20}
{\tilde{\cal H}} &=& \frac{1}{2}{\pi_{i}}M^i_j\pi^j + \frac 14 F_{ij}^2  + \frac{1}{2}\,m^2\,{A_{i}}^2 - m^2\partial^iA_i\theta +
\frac 12 m^2 \,(\partial_i\theta)^2,
\ea
whose the phase space metric is

\be
\label{matrix21}
M^i_j = g^i_j - \frac{\partial^i\partial_j}{m^2},
\ee
and has a first class constraint, $\chi_1$, which generates the infinitesimal transformations given in (\ref{Proca16}).

\subsection{The $O(N)$ invariant nonlinear sigma model} 

In this subsection, the hidden symmetry presents on the $O(N)$ nonlinear sigma model will be disclosed enlarging the phase space with the
introduction of WZ field {\it via} symplectic gauge-invariant formalism. We first apply the symplectic method to the original second class
model in order to show the second class nature of the model, and also to obtain the usual Dirac's brackets among the phase space fields.
Later, we unveil the hidden gauge symmetry of the model which dwells on the original phase space.

The $O(N)$ nonlinear sigma model in two dimensions is a free field theory for the multiplet $\sigma_a\equiv
(\sigma_1,\sigma_2,\dots,\sigma_n)$ satisfying a nonlinear constraint $\sigma_a^2=1$. This model has its dynamics governed by the
Lagrangian density

\begin{equation}
{\cal L}=\frac{1}{2}\,\partial_\mu\sigma^a\partial^\mu\sigma_a
- \frac{1}{2}\,\lambda\,\bigl(\sigma^a\sigma_a - 1\bigr),
\label{3001}
\end{equation}
where the $\mu=0,1$ and $``a"$ is an index related to the $O(N)$ symmetry group.  

In order to implement the symplectic method, the original second-order Lagrangian in the velocity, given in (\ref{3001}), is reduced into
a first-order form, given by,
 
\begin{equation}
\label{3002}
{\cal L}^{(0)} = \pi_a\dot{\sigma}^a - V^{0},
\end{equation}
with

\be
\label{3002a}
V^{(0)} = \frac{1}{2}\,\pi^2_a + \frac{1}{2}\,\lambda\,\bigl(\sigma^2_a - 1\bigr) - \frac{1}{2}\,{\sigma^\prime_a} ^2,
\ee
where dot and prime represent temporal and spatial derivatives, respectively. The symplectic fields are
$\xi_\alpha^{(0)}=(\sigma_a,\pi_a,\lambda)$ and the index ${(0)}$ indicates the zeroth-iterative. After, the symplectic tensor given in
Eq.(\ref{2010}) is computed as 

\begin{equation}
\label{3003}
f^{(0)} = \left(
\begin{array}{ccc}
0           & -\delta_{ab} & 0 \\
\delta_{ba} &         0     & 0 \\
0           &         0     & 0
\end{array}
\right)\,\delta(x-y).
\end{equation}
This matrix is singular, thus, it has a zero-mode, reads as

\begin{equation}
\label{3004}
\nu^{(0)} = \left(
\begin{array}{ccc}
{\bf 0} \\
{\bf 0} \\
1
\end{array}
\right).
\end{equation}
Contracting this zero-mode with the gradient of symplectic potential $V^{(0)}$, given in Eq.(\ref{3002a}), the following constraint is
obtained,

\be
\label{3005a}
\Omega_1 = \sigma^2_a- 1.
\ee
In agreement with the symplectic formalism, this constraint must be introduced into the canonical sector of the first-order Lagrangian
(\ref{3002}) through a Lagrange multiplier $\rho$, then, we get the first-iterative Lagrangian as

\begin{equation}
\label{3005}
{\cal L}^{(1)} = \pi_a\dot{\sigma}^a + \Omega_1\dot{\rho} - V^{(1)}\mid_{\Omega_1=0},
\ee
with

\be
\label{3006}
V^{1}\mid_{\Omega_1=0} =\frac{1}{2}\,\pi^2_a  + \frac{1}{2}\,{\sigma^\prime}^2_a.
\end{equation}
Now, the symplectic fields are $\xi_\alpha^{(1)}=(\sigma_a,\pi_a,\rho)$ with the following one-form canonical momenta,

\begin{eqnarray}
\label{3007}
A_{\sigma_a}^{(1)} &=& \pi_a, \nonumber \\
A_{\pi_a}^{(1)} &=& 0,  \\
A_{\rho}^{(1)} &=& \bigl(\sigma^2_a - 1\bigr).\nonumber
\end{eqnarray}
The corresponding symplectic tensor $f^{(1)}$, given by,  

\begin{equation}
\label{3008}
f^{(1)}=\left(
\begin{array}{ccc}
0           & -\delta_{ab} & \sigma_a \\
\delta_{ab} &         0     & 0 \\
-\sigma_b           &         0     & 0
\end{array}
\right)\,\delta(x-y),
\end{equation}
is singular, thus, it has a zero-mode that generates a new constraint, reads as 

\be
\label{3009}
\Omega_2 = \sigma_a\pi^a.
\ee

Introducing the constraint $\Omega_2$ into the first-iterated Lagrangian (\ref{3005})through a Lagrange multiplier $\zeta$, the
twice-iterated Lagrangian is obtained as

\begin{equation}
\label{3010}
{\cal L}^{(2)} = \pi_a \dot{\sigma}^a + \dot{\rho}\bigl(\sigma^2_a - 1\bigr)+ \dot{\zeta}(\sigma_a\pi^a) - V^{(2)},
\end{equation}
with $V^{(2)}$ = $V^{(1)}\mid_{\Omega_1=0}$. The enlarged symplectic fields are $\xi_\alpha^{(2)}=(\sigma_a,\pi_a,\rho,\zeta)$ and the
new one-form canonical momenta are

\begin{eqnarray}
\label{formula22}
A_{\sigma_a}^{(2)} &=& \pi_a, \nonumber \\
A_{\pi_a}^{(2)} &=& 0, \nonumber \\
A_{\rho}^{(2)} &=& \sigma^2_a - 1,\nonumber \\
A_{\zeta}^{(2)} &=& \sigma_a \pi^a. \nonumber
\end{eqnarray}
The corresponding matrix $f^{(2)}$ is

\begin{equation}
\label{3011}
f^{(2)}=\left(
\begin{array}{cccc}
0           & -\delta_{ab}  &  \sigma_a   &   \pi_a \\
\delta_{ba} &         0     &   0    &    \sigma_a \\
-\sigma_b         &         0     &   0    &     0   \\
-\pi_b       &        -\sigma_b    &   0    &     0  
\end{array}
\right)\,\delta(x-y),
\end{equation}
that is a nonsingular matrix. The inverse of $f^{(2)}$ gives the usual Dirac brackets among the physical fields, given by,

\ba
\lbrace \sigma_a(x),\sigma_b(y)\rbrace^* = \lbrace \pi_a(x),\pi_b(y)\rbrace^* &=& 0,\nonumber\\
\lbrace \sigma_a(x),\pi_b(y)\rbrace^* &=&\(\delta_{ab} - \frac{\sigma_a\sigma_b}{\sigma^2}\)\delta(x - y),\\
\lbrace \pi_a(x),\pi_b(y)\rbrace^* &=& \frac{(\sigma_a\pi_b - \sigma_b\pi_a)}{\sigma^2}\delta(x - y).\nonumber
\ea
This means that the NLSM is not a gauge invariant theory.

At this stage we are ready to implement our proposal. In order to disclose the hidden symmetry present on the NLSM, the original phase
space will be extended with the introduction of WZ field following the symplectic gauge-invariant formalism. This
process is based on the introduction of two arbitrary functions, $\Psi(\sigma_a,\pi_a,\theta)$ and  $G(\sigma_a,\pi_a,\theta)$, into the
first-order Lagrangian as follows,

\be
\label{3012a}
{\tilde {\cal L}}^{(0)} = \pi_a\dot{\sigma}^a + \Psi\dot{\theta} - {\tilde V}^{(0)},
\ee
where the symplectic potential is

\be
\label{3013a}
{\tilde V}^{(0)} = \frac{1}{2}\,\pi^2_a + \frac{1}{2}\,\lambda\,\bigl(\sigma^2_a - 1\bigr) + \frac{1}{2}\,{\sigma^\prime_a} ^2 + G(\sigma_a,\pi_a,\theta),
\ee
with $G(\sigma_a,\pi_a,\theta)$ satisfying the relations given in Eqs.(\ref{2060}) and (\ref{2070}).

The symplectic fields are $\tilde \xi_\alpha^{(0)}=(\sigma_a,\pi_a,\lambda,\theta)$
with the following one-form canonical momenta,

\begin{eqnarray}
\label{3015}
\tilde A_{\sigma_a}^{(0)} &=& \pi_a, \nonumber \\
\tilde A_{\pi_a}^{(0)} &=& 0, \nonumber \\
\tilde A_{\lambda}^{(0)} &=& \frac 12 (\sigma^2_a - 1), \nonumber \\
\tilde A_{\theta}^{(0)} &=& 0.
\end{eqnarray}
As established by the symplectic gauge-invariant formalism, the corresponding matrix $\tilde f^{(0)}$, given by

\begin{equation}
\label{3016}
\tilde f^{(0)} = \pmatrix{ 0  & -\delta_{ab}  &  0  &   \frac{\partial\Psi_y}{\partial\sigma^{x}_a} \cr
\delta_{ba} &  0 & 0  & \frac{\partial\Psi_y}{\partial\pi^{x}_a} \cr
0 & 0 & 0  & \frac{\partial\Psi_y}{\partial\lambda^{x}}  \cr
- \frac{\partial\Psi_x}{\partial\sigma^{y}_b} & - \frac{\partial\Psi_x}{\partial\pi^{y}_b}  & -\frac{\partial\Psi_x}{\partial\lambda^{y}} & f_{\theta_x\theta_y} }\delta(x-y),
\end{equation}
must be singular, that fixes the dependence relations of arbitrary function $\Psi$, namely,
$\frac{\partial\Psi_y}{\partial\lambda^{x}_a}=0$, {\it i.e}, $\Psi\equiv\Psi(\sigma_a,\pi_a,\theta)$. Subsequently, this matrix has a
zero-mode, identified as being the gauge symmetry generator. To pull out the hidden symmetry this zero-mode must to satisfy the relation
(\ref{matrix02}), allowing then the computation of $\Psi$. Let us to start considering the symmetry generated by the following zero-mode,

\begin{equation}
\label{3017}
\nu^{(0)}=\left(
\begin{array}{ccc}
{\bf 0} \\
\sigma_a \\
0 \\
1
\end{array}
\right).
\end{equation}
Since this zero-mode and the symplectic matrix (\ref{3016}) satisfy the relation (\ref{matrix02}), $\Psi$ is determined as

\be
\label{3017a}
\Psi = \sigma_a^2 + c,
\ee
where ``c" is a constant parameter. This completes the first step of the symplectic gauge-invariant formalism.

Now, the second step starts. Imposing that no more constraint is generated by the contraction of the zero-mode with the gradient of
potential, the correction terms in order of $\theta$ can be explicitly computed. The first-order correction term in
$\theta$, ${\cal G}^{(1)}$, determined after an integration process, is

\begin{equation}
\label{3018}
{\cal G}^{(1)}(\sigma_a,\pi_a,\theta) = - \sigma_a\pi_a\theta.
\end{equation}
Bringing back this expression into the Eq.(\ref{3013a}), the new Lagrangian is obtained as

\begin{equation}
\label{3019}
\tilde {\cal L}^{(0)} = \pi_a\dot{\sigma}^a + \Psi\dot{\theta} - \frac{1}{2}\,{\sigma^\prime} ^2_a - \frac{1}{2}\,\pi^2_a - \frac{1}{2}\,\lambda\,\bigl(\sigma^2_a - 1\bigr)  + \sigma_a\pi_a\theta.
\end{equation}

However, the model is not yet a gauge invariant because the contraction of the zero-mode $\nu^{(0)}$ with the gradient of potential
$V^{0}$ produces a non null value, indicating that it is necessary to compute the remaining correction terms ${\cal G}^{(n)}$ in order of
$\theta$. It is achieved just imposing that the zero-mode does not generate a new constraint. It allows us to determine the second-order
correction term ${\cal G}^{(2)}$, given by

\begin{eqnarray}
\label{3020}
{\cal G}^{(2)} = + \frac {1}{2} \sigma_a^2 \theta^2.
\end{eqnarray}
Bringing this result into the first-order Lagrangian (\ref{3019}), we obtain

\begin{equation}
\label{3021}
\tilde {\cal L}^{(0)} = \pi_a\dot{\sigma}^a + \Psi\dot{\theta} - \frac{1}{2}\,{\sigma^\prime} ^2_a -
\frac{1}{2}\,\lambda\,\bigl(\sigma^2_a - 1\bigr)  - \frac{1}{2}\,\pi^2_a + \sigma_a\pi^a\theta - \frac {1}{2} \sigma_a^2 \theta^2.
\end{equation}
Now the zero-mode $\nu^{(0)}$ does not produce a new constraint, consequently, the model has a symmetry and, in accordance with the
symplectic point of view, the generator of the symmetry is the zero-mode. Due to this, all correction terms ${\cal G}^{(n)}$ with
$n\geq 3$ are null.

At this moment, we are interested to recover the invariant second-order Lagrangian from its first-order form given in Eq.(\ref{3021}).
To this end, the canonical momenta must be eliminated from the Lagrangian (\ref{3021}). From the equation of motion for $\pi_a$, the
canonical momenta are computed as

\begin{equation}
\label{3022}
\pi_a = \dot \sigma_a + \sigma_a\theta.
\end{equation}
Inserting this result into the first-order Lagrangian (\ref{3021}), we get the second-order Lagrangian as

\begin{equation}
\label{3023}
\tilde {\cal L} = \frac{1}{2}\,\partial_\mu\sigma_a\partial^\mu\sigma^a - (\dot{\sigma_a}\sigma^a)\theta -
\frac{1}{2}\bigl(\sigma^2_a - 1\bigr)\lambda ,
\end{equation}
with the following gauge invariant Hamiltonian,

\begin{equation}
\label{3024}
\tilde {\cal H} = \frac{1}{2}\,\pi^2_a  + \frac{1}{2}\,{\sigma^\prime} ^2_a - (\sigma_a\pi^a)\theta +
\frac{1}{2}\lambda \bigl(\sigma^2_a - 1\bigr)+ \frac{1}{2}\,\sigma^2_a\theta^2 .
\end{equation}
By construction, both Lagrangian (\ref{3023}) and Hamiltonian (\ref{3024}) are gauge invariant. From the Dirac point of view,
$\Omega_1$ arises as a secondary constraint from the temporal stability imposed on the primary constraints, $\pi_\lambda$ and
$\pi_\theta$,  and plays the role of the Gauss law, which generates the time independent gauge transformation. To proceed the
quantization, we recognize the states of physical interest as those that are annihilated by $\Omega_1$. This gauge invariant formulation
of the NLSM was also obtained by one of us in \cite{WN} with the introduction of WZ fields, as established by iterative method\cite{IJMP},
and by another authors using the BFFT formalism\cite{BGB}.

To become this work self-consistent the infinitesimal gauge transformation will be determined. As established by the symplectic formalism,
the zero-mode is identified as being the generator of the infinitesimal gauge transformations
$\delta\tilde \xi_\alpha^{(0)}=\varepsilon \nu^{(0)}$, namely,

\begin{eqnarray}
\label{3025}
\delta \sigma_a &=& 0,\nonumber\\
\delta \pi_a &=& \varepsilon \sigma_a ,\nonumber\\
\delta \lambda &=& 0,\\
\delta \theta &=& \varepsilon.\nonumber
\end{eqnarray}
For the transformation above the Hamiltonian changes as

\be
\label{3025a}
\delta{\cal H} = 0.
\ee

Similar results were also obtained in the literature using different methods based on the Dirac's constraint
framework\cite{WN,BGB,JW1,JW2,HKP,NW}. However, these techniques are affected by some ambiguities problems that naturally arise when the
second class nature of the set of constraints transmutes to first class with the introduction of the WZ fields. In our procedure, this
kind of problem does not arise, consequently, the arbitrariness disappears.

Henceforth, we are interested to disclose the hidden symmetry of the NLSM lying on the original phase space $(\sigma_a,\pi_a)$. To this
end, we use the Dirac method to obtain the set of constraints of the gauge invariant NLSM described by the Lagrangian (\ref{3023}) and
Hamiltonian (\ref{3024}), given by,

\begin{eqnarray}
\label{3026}
\phi_1 &=& \pi_\lambda,\nonumber\\
\phi_2 &=& - \frac 12(\sigma^2_a - 1),
\end{eqnarray}
and

\begin{eqnarray}
\label{3027}
\varphi_1 &=& \pi_\theta,\nonumber\\
\varphi_2 &=& \sigma_a\pi_a - \sigma_a^2\theta,
\end{eqnarray}
where $\pi_\lambda$ and $\pi_\theta$ are the canonical momenta conjugated to $\lambda$ and $\theta$, respectively. The corresponding Dirac
matrix is singular, however, there are nonvanishing Poisson brackets among some constraints, indicating that there are both second class
and first class constraints. It is solved splitting up the second class constraints from the first class ones through constraint
combination. The set of first class constraints is

\begin{eqnarray}
\label{3028}
\chi_1 &=& \pi_\lambda,\nonumber\\
\chi_2 &=& - \frac 12 (\sigma^2_a - 1) + \pi_\theta,
\end{eqnarray}
while the set of second class constraints is

\begin{eqnarray}
\label{3029}
\varsigma_1 &=& \pi_\theta,\nonumber\\
\varsigma_2 &=& \sigma_a\pi_a - \sigma^2_a\theta .
\end{eqnarray}
Since the second class constraints are assumed equal to zero in a strong way, and using the Maskawa-Nakajima theorem\cite{NM}, the Dirac's
brackets are worked out as 

\begin{eqnarray}
\label{3030}
\lbrace \sigma_i(x), \sigma_j(y)\rbrace^* &=& 0,\nonumber\\
\lbrace \sigma_i(x), \pi_j(y)\rbrace^* &=& \delta_{ij}\,\delta(x-y),\\
\lbrace \pi_i(x), \pi_j(y)\rbrace^* &=& 0.\nonumber
\end{eqnarray}
Hence, the gauge invariant Hamiltonian is rewritten as

\begin{eqnarray}
\label{3031}
\tilde {\cal H} &=& \frac{1}{2}\pi^2_a + \frac{1}{2}\,{\sigma^\prime} ^2_a 
- \frac{1}{2}\frac{(\sigma_a\pi^a)^2}{\sigma_a\sigma^a} + \frac{\lambda}{2}(\sigma^2_a -1)\nonumber\\
&=& {1\over 2} \pi_iM_{ij}\pi_j + \frac{1}{2}\,{\sigma^\prime}^2_a + \frac{\lambda}{2}(\sigma_a^2 - 1),
\end{eqnarray}
where the phase space metric $M_{ij}$, given by

\be
\label{3032}
M_{ij} = \delta_{ij} - \frac {\sigma_i\sigma_j}{\sigma_k^2},
\ee
is a singular matrix, and the set of first class becomes

\ba
\label{3033}
\chi_1 &=& \pi_\lambda,\nonumber\\
\chi_2 &=& - \frac 12(\sigma^2_a - 1).
\ea
Note that the constraint $\chi_2$, originally a second class constraint, becomes the generator of the gauge symmetry, satisfying the
first class property

\begin{equation}
\label{3035}
\{\chi_2, \tilde{H} \} = 0.
\end{equation}
Due to this, the infinitesimal gauge transformations are computed as

\ba
\label{3036}
\delta \sigma_a &=& \varepsilon\lbrace \sigma_a,\chi_2\rbrace = 0,\nonumber\\
\delta \pi_a &=& \varepsilon\lbrace\pi_a,\chi_2\rbrace=\varepsilon \sigma_a,\\
\delta\lambda &=& 0.\nonumber
\ea
where $\varepsilon$ is an infinitesimal parameter. It is easy to verify that the Hamiltonian (\ref{3031}) is invariant under these
transformations because $\sigma_a$ are eigenvectors of the phase space metric ($M_{ij}$) with null eigenvalues. In this section we
reproduce the results originally obtained in \cite{KR} from an alternative point of view.

\subsection{The gauge invariant bosonized CSM}

It has been shown over the last decade that anomalous gauge theories in two dimensions can be consistently and unitarily quantized for
both Abelian\cite{RR,JR,many} and non-Abelian\cite{RR1,LR} cases. In this scenario, the two dimensional model that has been extensively
studied is the CSM. We start with the following Lagrangian density of the bosonized CSM with $a > 1$,

\be
{\cal L} = -\frac 14 \, F_{\mu\nu}\,F^{\mu\nu} + 
            \frac{1}{2}\,\partial_\mu\phi\,\partial^{\mu}\phi +
            q\,\left(g^{\mu\nu} -\,\epsilon^{\mu\nu}\right)\,
            \partial_{\mu}\phi\,A_{\nu} +
           \frac{1}{2}\,q^2a\,A_{\mu}A^{\mu}\,.
\label{00000}
\ee
Here, $F_{\mu\nu} = \partial_{\mu}A_{\nu}-\partial_{\nu}A_{\mu}$, $g_{\mu\nu} =
\mbox{diag}(+1,-1)$ and $\epsilon^{01} = -\epsilon^{10} = \epsilon_{10} = 1$. Afterhere, the symplectic method will be used to quantize
the original second class model and, then, the Dirac's brackets and the respective reduced Hamiltonian will be determined as well.
In order to implement the symplectic method, the original second-order Lagrangian in the velocity, given in (\ref{00000}), is reduced
into its first-order as follows,
\be
\label{mitra1}
L^{(0)} =\pi _\phi \dot{\phi} + \pi _1 \dot{A_1} - U^{(0)},
\end{equation}
where the zeroth-iterative symplectic potential $U^{(0)}$ is

\begin{eqnarray}
\label{mitra2}
U^{(0)}&=& {1\over 2}(\pi _1^2 +\pi _\phi ^2 +\phi ^{\prime 2}) - A_0( \pi _1^\prime +
{1\over2}q^2(a-1)A_0 + q^2A_1 + q\pi _\phi + q\phi^\prime )\nonumber \\ 
&-& A_1 (-q\pi_\phi -{1\over 2}q^2(a+1)A_1 - q\phi^\prime ),
\end{eqnarray}
where dot and prime represent temporal and spatial derivatives, respectively. The zeroth-iterative symplectic variables are
$\xi _\alpha^{(0)}=( \phi ,\pi _\phi,A_0 ,A_1 ,\pi _1)$ with
the following one-form canonical momenta $A_\alpha $,

\begin{eqnarray}
\label{00050}
A_\phi ^{(0)} &=& \pi _\phi, \nonumber \\
A_{A_1}^{(0)} &=& \pi_1, \\
A_{\pi _\phi}^{(0)}=A_{A_0}^{(0)}=A_{\pi _1}^{(0)} &=& 0.  \nonumber
\end{eqnarray}
the zeroth-iterative symplectic tensor is obtained as

\begin{equation}
\label{00060}
f^{(0)}(x,y)= \left( \begin{array}{ccccc}
0 & -1 & 0 & 0 & 0 \\
1 & 0 & 0 & 0 & 0 \\
0 & 0 & 0 & 0 & 0 \\
0 & 0 & 0 & 0 & -1 \\
0 & 0 & 0 & 1 & 0 
\end{array} \right )\delta(x - y).
\end{equation}
This matrix is obviously singular, thus, it has a zero-mode that generates a constraint when contracted with the gradient of the
potential $U^{(0)}$, given by,

\begin{eqnarray}
\label{00080}
\Omega _1 &=& \nu_\alpha ^{(0)}{{\partial U^{(0)}}\over {\partial \xi _\alpha ^{(0)}}}\nonumber \\
&=&\pi _1^\prime +q^2(a-1)A_0 + q^2A_1 + q\pi _\phi + q\phi ^\prime,
\end{eqnarray}
that is identified as being the Gauss law, which satisfies the following Poisson algebra,

\begin{equation}
\label{00081}
\lbrace\Omega _1(x),\Omega _1(y)\rbrace = 0. 
\end{equation}
Bringing back the constraint $\Omega _1$ into the canonical sector of the
first-order Lagrangian through a Lagrange multiplier $\eta $, we get the first-iterative Lagrangian $L^{(1)}$, namely,

\begin{equation}
\label{00090}
L^{(1)} =\pi _\phi \dot{\phi} + \pi _1 \dot{A_1} + \Omega_1 \dot {\eta } - U^{(1)},
\end{equation}
with the first-order symplectic potential

\begin{eqnarray}
\label{00100}
U^{(1)}&=& {1\over 2}(\pi _1^2 +\pi _\phi ^2 +\phi ^{\prime 2}) - A_0( \pi _1^\prime +
{1\over2}q^2(a-1)A_0 + q^2A_1 + q\pi _\phi + q\phi^\prime )\nonumber \\ 
&-& A_1 (-q\pi_\phi -{1\over 2}q^2(a+1)A_1 - q\phi^\prime ),
\end{eqnarray}
where $U^{(1)}=U^{(0)}$. Therefore, the symplectic variables become $\xi _\alpha^{(1)}=( \phi ,\pi _\phi, A_0 ,A_1 ,\pi _1, \eta )$
with the following one-form canonical momenta,

\begin{eqnarray}
\label{00110}
A_\phi ^{(1)} &=& \pi _\phi ,\nonumber \\
A_{A_1}^{(1)} &=& \pi _1 ,\nonumber \\
A_{A_0}^{(1)} = A_{\pi _\phi}^{(1)}= A_{\pi _1}^{(1)}&=& 0, \\
A_{\eta}^{(1)}&=& \pi _1^\prime +q^2(a-1)A_0 + q^2A_1 + q\pi _\phi + q\phi ^\prime. \nonumber
\end{eqnarray}
The corresponding matrix $f^{(1)}$ is then

\begin{equation}
\label{00120}
f^{(1)}(x, y)= \left ( \begin{array}{cccccc}
0 & -1 & 0 & 0 & 0 & q\partial_y \\
1 & 0 & 0 & 0 & 0 & q \\
0 & 0 & 0 & 0 & 0 & q^2(a-1) \\
0 & 0 & 0 & 0 & -1 & q^2 \\
0 & 0 & 0 & 1 & 0 & \partial_y \\
-q\partial _x & - q & - q^2(a-1) & -q^2 & - \partial_x & 0 
\end{array} \right )\delta (x-y),
\end{equation}
that is a nonsingular matrix. The inverse of $f^{(1)}(x,y)$ gives, after a straightforward calculation, the Dirac brackets among
the physical fields, read as

\begin{eqnarray}
\label{00180}
\left\{\,\phi(x)\,,\,\phi(y)\,\right\}^{*}&=& 0, \nonumber \\
\left\{\,\phi(x)\,,\,\pi_\phi(y)\,\right\}^{*}&=& \delta (x - y)\, , \nonumber \\
\left\{\,\phi(x)\,,\, A_0(y)\,\right\}^{*}&=& -\frac{1}{q(a-1)}\,
                                       \delta (x - y)\, , \nonumber \\
\left\{\,\phi(x)\,,\, A_1(y)\,\right\}^{*}&=& 0, \nonumber \\
\left\{\,\phi(x)\,,\,\pi_1(y)\,\right\}^{*}&=& 0, \nonumber \\
\left\{\pi_\phi(x),\pi_\phi(y)\right\}^{*} &=& 0, \nonumber \\
\left\{\pi_\phi(x),A_0(y)\right\}^{*} &=& \frac{1}{q(a-1)}
                                       \partial_y \delta (x - y)\, , \nonumber \\
\left\{\pi_\phi(x),A_1(y)\right\}^{*} &=& 0, \nonumber \\
\left\{\pi_\phi(x),\pi_1(y)\right\}^{*} &=& 0, \nonumber \\
\left\{A_{1}(x),A_{0}(y)\right\}^{*} &=& -\frac{1}{q^{2}(a-1)}\, 
                                        \partial_y\delta (x - y)\, , \\
\left\{A_{1}(x),A_{1}(y)\right\}^{*} &=& 0, \nonumber\\  
\left\{A_{1}(x),\pi_{1}(y)\right\}^{*} &=& \delta (x - y)\, ,\nonumber \\
\left\{\pi_{1}(x),A_0(y)\right\}^{*} &=& \frac{1}{(a-1)}
                                           \delta (x - y)\,\,, \nonumber\\ 
\left\{\pi_{1}(x),\pi_{1}(y)\right\}^{*} &=& 0. \nonumber   
\end{eqnarray}
This means that the model is not a gauge invariant theory.

Afterhere, the gauge symmetry present on the model will be disclosed via a new gauge-invariant formalism that not require more than one
WZ field. The fundamental concept behind the symplectic gauge-invariant formalism dwells on the extension of the original phase space with
the introduction of two arbitrary function $\Psi(\phi,\pi_\phi,A_0,A_1,\pi_1,\theta)$ and $G(\phi,\pi_\phi,A_0,A_1,\pi_1,\theta)$,
depending on the original phase space
variables and the WZ variable $\theta$, into the first-order Lagrangian, right on the kinetical and symplectic potential sector,
respectively. In views of this, the first-order Lagrangian that
governs the dynamics of the bosonized CSM, given in Eq.(\ref{mitra1}), is rewritten as

\begin{equation}
\label{00300}
{\tilde L}^{(0)} =\pi _\phi \dot{\phi} + \pi _1 \dot{A_1} + \dot\theta\Psi - {\tilde U}^{(0)},
\end{equation}
where

\begin{eqnarray}
\label{00310}
{\tilde U}^{(0)}&=& {1\over 2}(\pi _1^2 +\pi _\phi ^2 +\phi ^{\prime 2}) - A_0( \pi _1^\prime +
{1\over2}q^2(a-1)A_0 + q^2A_1 + q\pi _\phi + q\phi^\prime )\nonumber \\ 
&-& A_1 (-q\pi_\phi -{1\over 2}q^2(a+1)A_1 - q\phi^\prime ) + G(\phi , \pi _\phi, A_0, A_1, \pi _1, \theta ).
\end{eqnarray}

The gauge-invariant formulation comprehend two steps; one is dedicated to the computation of $\Psi$ while the other is addressed to the
calculation of $G$. 

The enlarged symplectic variables are now ${\tilde \xi} _\alpha^{(0)}=( \phi ,\pi _\phi, A_0 ,A_1 ,\pi _1, \theta )$ with the following
one-form canonical momenta

\begin{eqnarray}
\label{00320}
{\tilde A}_\phi ^{(0)} &=& \pi _\phi ,\nonumber \\
{\tilde A}_{A_1}^{(0)} &=& \pi _1 ,\nonumber \\
{\tilde A}_{A_0}^{(0)} = {\tilde A}_{\pi _\phi}^{(0)}={\tilde A}_{\pi _1}^{(0)} &=& 0, \\
{\tilde A}_{\theta}^{(0)} &=& \Psi \nonumber.
\end{eqnarray}
The corresponding symplectic matrix ${\tilde f}^{(0)}$ reads

\begin{equation}
\label{00325}
{\tilde f}^{(0)}= \pmatrix{0 & -1 & 0 & 0 & 0 & \frac{\partial\Psi^y}{\partial \phi^x} \cr
1 & 0 & 0 & 0 & 0 & \frac{\partial\Psi^y}{\partial \pi^x_\phi}\cr
0 & 0 & 0 & 0 & 0 & \frac{\partial\Psi^y}{\partial A^x_0}\cr
0 & 0 & 0 & 0 & -1 & \frac{\partial\Psi^y}{\partial A^x_1}\cr
0 & 0 & 0 & 1 & 0 & \frac{\partial\Psi^y}{\partial \pi^x_1}\cr
- \frac{\partial\Psi^x}{\partial \phi^y} & - \frac{\partial\Psi^x}{\partial \pi^y_\phi}& -\frac{\partial\Psi^x}{\partial A^y_0} & 
- \frac{\partial\Psi^x}{\partial A^y_1} & -
\frac{\partial\Psi^x}{\partial \pi^y_1}  & f_{\theta_x\theta_y}} \delta (x-y),
\end{equation}
where

\be
\label{00325a}
f_{\theta_x\theta_y} = \frac{\partial \Psi_y}{\partial \theta_x} - \frac{\partial \Psi_x}{\partial \theta_y},
\ee
with $\theta_x \equiv \theta(x)$, $\theta_y \equiv \theta(y)$, $\Psi_x \equiv \Psi(x)$ and $\Psi_y \equiv \Psi(y)$. Note that this matrix
is singular since $\frac{\partial\Psi^x}{\partial A^y_0} = 0$. Due to this, we conclude that
$\Psi\equiv\Psi(\phi,\pi_\phi,A_1,\pi_1,\theta)$.

To unveil the gauge symmetry presents on the model, we assume that this singular matrix has a zero-mode $(\nu^{(0)})$ that satisfies the
following relation,

\begin{equation}
\label{00326}
\int \nu^{(0)}_\alpha (x) {\tilde f}^{(0)}_{\alpha\beta}(x - y)\; d\,y = 0.
\end{equation}
>From this relation a set of equations will be obtained and, consequently, the arbitrary function $\Psi$ can be determined. We consider to
study the symmetry related with the following zero-mode,

\begin{equation}
\label{00327}
\bar\nu^{(0)} = \pmatrix{  q & - q\partial_x & 1 & \partial_x & - q^2  & - 1},
\end{equation}
with bar representing a transpose matrix.

To start, we contract the zero-mode (\ref{00327}) with the symplectic matrix (\ref{00325}), as shown in Eq.(\ref{00326}). Due to this,
some equations arise and, after an integration process, $\Psi$ is determined as

\begin{equation}
\label{00327a}
\Psi = \pi^\prime + q\phi^\prime + q\pi_\phi + q^2A_1.
\end{equation}
The corresponding symplectic matrix (\ref{00325}) is rewritten as
\begin{equation}
\label{00330}
{\tilde f}^{(0)}= \left ( \begin{array}{cccccc}
0 & -1 & 0 & 0 & 0 &  q\partial_y\\
1 & 0 & 0 & 0 & 0 & q\\
0 & 0 & 0 & 0 & 0 & 0\\
0 & 0 & 0 & 0 & -1 & q^2 \\
0 & 0 & 0 & 1 & 0 & \partial _y\\
- q\partial_x & - q & 0 & - q^2 & - \partial_x & 0  
\end{array} \right )\delta(x-y)
\end{equation}
that is obviously singular, consequently, has a zero-mode that, by construction, is given in Eq.(\ref{00327}).

Afterhere, we start the second step to reformulate the  model as a gauge invariant theory. At this stage, the correction terms in order of
$\theta$, embraced by the arbitrary function $G$, given in Eq.(\ref{2060}), will be computed. It is achieved just imposing that no more
constraint arises from the contraction of the zero-mode, given in Eq.(\ref{00327}), with the gradient of the symplectic potential,

\begin{equation}
\label{00331}
\nu^{(0)}_\alpha\frac{\partial {\tilde U}^{(0)}}{\partial \xi^{(0)}_\alpha} = 0.
\end{equation}

The first-order correction term in $\theta $, ${\cal G}^{(1)}$, is determined as

\be
\label{00350}
{\cal G}^{(1)}(\phi , \pi_\phi, A_1, \pi_1, A_0, \theta ) = - \Omega_1 \theta + q^2(a-1)A^{\prime}_1\theta -
q^2\theta \pi _1,
\ee
after an integration process. Bringing back this expression into the Eq. (\ref{00300}), the new Lagrangian is obtained as

\begin{equation}
\label{00360}
{\tilde L}^{(0)} = \pi _\phi \dot{\phi} + \pi _1 \dot{A_1} + \Psi\dot {\theta } - {\tilde U}^{(0)},
\end{equation}
with

\begin{eqnarray}
\label{00361}
{\tilde U}^{(0)}&=& {1\over 2}(\pi _1^2 +\pi _\phi ^2 +\phi ^{\prime 2}) - A_0( \pi _1^\prime +{1\over2}q^2(a-1)A_0 + q^2A_1 + q\pi _\phi + q\phi^\prime )\nonumber \\ 
&-& A_1 (-q\pi_\phi -{1\over 2}q^2(a+1)A_1 - q\phi^\prime ) 
-\Omega_1 \theta + q^2(a-1)\theta^\prime A_1 - q^2\theta \pi _1.
\end{eqnarray}
That it is not yet a gauge invariant Lagrangian, because the zero-mode $\bar\nu^{(0)}$ still generates a new constraint, given by

\begin{equation}
\label{00370}
\nu_\alpha^{(1)}{{\partial {\tilde U}^{(0)}}\over {\partial \xi _\alpha ^{(0)}}} = q^2(a-1)\theta^{\prime\prime} -
q^2(a-1) \theta + q^4\theta, 
\end{equation}
indicating that it is necessary to obtain the remaining correction terms ${\cal G}^{(n)}$ in order of $\theta $. It is achieved just imposing that no more constraints are generated by the contraction of the zero-mode with the gradient of extended symplectic potential. It
allows us to determine the second-order correction term ${\cal G}^{(2)}$, reads as

\begin{eqnarray}
\label{00380}
\nu_\alpha^{(0)}{{\partial {\tilde U}^{(0)}}\over {\partial \xi _\alpha ^{(0)}}}&=& - q^2(a-1)\theta + q^2(a-1)\theta^{\prime\prime} +
q^4\theta - {{\partial {\cal G}^{(2)}}\over {\partial \theta }}= 0, \nonumber \\
{\cal G}^{(2)} &=& - {1\over 2}\;\; q^2(a-1){\theta ^{\prime}}^2 + {1\over 2}q^4{\theta }^2 - {1\over 2}q^2(a-1)\theta ^2.
\end{eqnarray}

Hence, the first-order Lagrangian (\ref{00360}) becomes

\begin{equation}
\label{00390}
{\tilde L}^{(0)} = \pi _\phi \dot{\phi} + \pi _1 \dot{A_1} + \Psi\dot\theta - {\tilde U}^{(0)},
\end{equation}
with the new symplectic potential

\begin{eqnarray}
\label{00391}
{\tilde U}^{(0)} &=& {1\over 2}(\pi _1^2 +\pi _\phi ^2 +\phi ^{\prime 2}) - A_0( \pi _1^\prime +{1\over2}q^2(a-1)A_0 + q^2A_1 +
q\pi _\phi + q\phi^\prime )\nonumber \\ 
&-& A_1 (-q\pi_\phi - {1\over 2}q^2(a+1)A_1 - q\phi^\prime ) - \Omega_1\theta + q^2(a-1)\theta A^\prime_1 - q^2\theta \pi _1 \nonumber \\
&-&{1\over 2}\;\; q^2(a-1){\theta ^{\prime}}^2 + {1\over 2}q^4{\theta }^2 - {1\over 2}q^2(a-1)\theta ^2.
\end{eqnarray}
The contraction of the zero-mode $\bar\nu^{(0)}$ with the new symplectic potential above does not produce a new constraint, consequently, the model has a symmetry and this zero-mode is the generator of the infinitesimal gauge transformation. Due to this, all correction
terms ${\cal G}^{(n)}$ with $n \geq 3$ are null. The infinitesimal gauge transformations generated by the zero-mode $(\delta\xi_i=\varepsilon \nu^{(0)})$ are

\ba
\label{00395}
\delta \phi &=& q\varepsilon,\nonumber\\
\delta \pi_\phi &=&  q \varepsilon^{\prime},\nonumber\\
\delta A_0 &=& \varepsilon,\nonumber\\
\delta A_1 &=& - \varepsilon^{\prime},\\
\delta \pi_1 &=& - q^2\varepsilon,\nonumber\\
\delta \theta &=& - \varepsilon.\nonumber
\ea
It is easy to verify that the Hamiltonian, identified as being the new symplectic potential ${\tilde U}^{(0)}$, is invariant under these infinitesimal gauge transformation above, namely,

\begin{equation}
\label{00396}
\delta {\tilde U}^{(0)} = 0.
\ee

At this point, we are interested to investigate this result and also to demonstrated that the anomaly was canceled. It will be done from the Dirac's point of view. From the Lagrangian (\ref{00390}) the chains of primary constraints are computed, namely,

\begin{eqnarray}
\label{00406}
\varphi_1 &=& \pi_0,\nonumber\\
\chi_1 &=& - \pi_\theta  + \Psi.
\end{eqnarray}
Afterward, these primary constraints are introduced into the Hamiltonian through Lagrange multipliers. In this way, the primary
Hamiltonian reads as

\be
\label{00407}
{\tilde U}^{(0)}_{primary} = {\tilde U}^{(0)} + \lambda_1\varphi_1 + \gamma_1\chi_1.
\ee
Since the constraint $\varphi_1$ has no time evolution, the following secondary constraint is required

\begin{equation}
\label{00407a1}
\varphi_2 =\Omega_1 - q^2(a-1)\theta,
\end{equation}
and no more constraints arise from the temporal stability condition. In this way, the total Hamiltonian is 

\be
\label{00407a3}
{\tilde U}^{(0)}_{total} = {\tilde U}^{(0)} + \lambda_1\varphi_1 + \lambda_2\varphi_2 + \gamma_1\chi_1.
\ee
The temporal stability condition for the constraint $\chi_1$ just allows to determine the Lagrange multiplier $\lambda_3$. In this way,
the gauge invariant version of the model has three constraints, namely, $\varphi_1$, $\varphi_2$ and $\chi_1$. The corresponding Dirac
matrix, given by,

\be
\label{00407a4}
C(x - y) = \left( \begin{array}{ccc}
0 & -q^2(a-1) & 0 \\
q^2(a-1) & 0 & q^2(a-1) \\
0 & -q^2(a-1) & 0 \\
\end{array}\right) \delta(x-y),
\ee  
is singular. As the Dirac matrix is singular, the model has both first class and second class constraints. Through a constraint
combination, we obtain a set of first class constraints, reads as

\be
\label{00408a}
\tilde\chi_1 = - \pi_{\theta}  + \Psi - \pi_0 ,
\ee
and a set of second class constraints, given by

\begin{eqnarray}
\label{00408}
\tilde\varphi_1 &=& \varphi_1,\nonumber\\
\tilde\varphi_2 &=& \Omega_1 - q^2(a-1)\theta .
\ea
It is easy to verify that $\tilde\chi_1$ is a first class constraint, identified as the Gauss law, while the remaining are second class constraints. Note that the anomaly was removed. Hence, the Gauss law is also recognized as being the generator of the gauge transformation given in Eq.(\ref{00395}).

At this stage, we will compute the degrees of freedom of the gauge invariant model proposed by us. The model has one first class and two second class constraints and the phase space dimensions sum eight dependent fields, i.e.,
$(\phi,\pi_\phi, A_1,\pi_1,A_0,\pi_0,\theta,\pi_\theta)$. The first class constraint eliminates two fields, while the second class constraints eliminate two fields, summing then four fields eliminated. Due to this, the model has four independent fields, i.e., there are two independent degrees of freedom.

In order to obtain the Dirac brackets, the set of second class constraints, $\tilde\varphi_1$ and $\tilde\varphi_2$ , will be assumed equal to zero in a strong way. After a straightforward computation, the Dirac's brackets among the phase space fields are obtained as
 
\ba
\label{00408a1}
\lbrace\phi(x),\phi(y)\rbrace^* &=& 0,\nonumber\\
\lbrace\phi(x),\pi_\phi(y)\rbrace^* &=& \delta(x - y),\nonumber\\
\lbrace\phi(x),A_0(y)\rbrace^* &=& - \frac{1}{q(a-1)}\delta(x - y),\nonumber\\
\lbrace\phi(x),A_1(y)\rbrace^* &=& 0,\nonumber\\
\lbrace\phi(x),\pi_1(y)\rbrace^* &=& 0,\nonumber\\
\lbrace\phi(x),\theta(y)\rbrace^* &=& 0,\nonumber\\
\lbrace\phi(x),\pi_\theta(y)\rbrace^* &=& 0,\nonumber\\
\lbrace\pi_\phi(x),A_0(y)\rbrace^* &=& \frac{1}{q(a-1)}\partial_y\delta(x - y),\nonumber\\
\lbrace\pi_\phi(x),\pi_\phi(y)\rbrace^* &=& 0,\nonumber\\
\lbrace\pi_\phi(x),A_1(y)\rbrace^* &=& 0,\nonumber\\
\lbrace\pi_\phi(x),\pi_1(y)\rbrace^* &=& 0,\nonumber\\
\lbrace\pi_\phi(x),\theta(y)\rbrace^* &=& 0,\\
\lbrace\pi_\phi(x),\pi_\theta(y)\rbrace^* &=& 0,\nonumber\\
\lbrace A_1(x),A_0(y)\rbrace^* &=& -\frac{1}{q^2(a-1)}\partial_y\delta(x - y),\nonumber\\
\lbrace A_1(x),A_1(y)\rbrace^* &=& 0,\nonumber\\
\lbrace A_1(x),\pi_{1}(y)\rbrace^* &=& \delta(x - y),\nonumber\\
\lbrace A_1(x),\theta(y)\rbrace^* &=& 0,\nonumber\\
\lbrace A_1(x),\pi_\theta(y)\rbrace^* &=& 0,\nonumber\\
\lbrace\pi_1(x),A_0(y)\rbrace^* &=&  \frac{1}{(a-1)}\delta(x - y),\nonumber\\
\lbrace\pi_1(x),\pi_1(y)\rbrace^* &=& 0,\nonumber\\
\lbrace\pi_1(x),\theta(y)\rbrace^* &=& 0, \nonumber\\
\lbrace\pi_1(x),\pi_\theta(y)\rbrace^* &=& 0, \nonumber\\
\lbrace\theta(x),A_0(y)\rbrace^* &=& 0, \nonumber\\
\lbrace\theta(x),\pi_\theta(y)\rbrace^* &=& \delta(x - y), \nonumber\\
\lbrace\pi_\theta(x),A_0(y)\rbrace^*  &=& \delta(x - y),\nonumber\\
\lbrace\pi_\theta(x),\pi_\theta(y)\rbrace^* &=& 0. \nonumber
\ea
Note that the Dirac brackets among the original phase space fields were obtained before.
After this process, the model passes to have one first class constraint only, reads as 

\be
\label{00409a1}
\chi=\tilde\chi_1|_{\tilde\varphi_1 = \tilde\varphi_2 = 0} = - \pi_\theta  + \Psi
\ee
identified as being the Gauss law, that satisfies the Poisson algebra, reads as

\be
\label{00409a2}
\lbrace\chi(x),\chi(y)\rbrace^* = 0.
\ee
In this way, the anomaly was eliminated, the symmetry was preserved, and the fundamental brackets among the original phase space fields were
reobtained. Note that the Gauss law is the generator of the gauge symmetry given in (\ref{00395}).

Once more, the counting of the independent degrees of freedom matches with the result obtained in the second class case. The invariant
model has a phase space $(\phi,\pi_\phi,A_1,\pi_1,\theta,\pi_\theta)$, summing six dependent fields, and has a first class constraint
which eliminates two fields, consequently, the model has two independent degrees of freedom.

At this point, we are interested to comment some consistency for the gauge invariant version of the bosonized CSM. To this end, the
remaining symmetry will be eliminated with the introduction of the unitary gauge-fixing term, given by,

\be
\label{00409a3}
\theta = 0.
\ee
Due to this, both noninvariant Hamiltonian and the corresponding Dirac brackets computed in the beginning of this section are reobtained,
resuscitating then the anomaly. In views of this,
we conclude that the new symplectic gauge-invariant formalism does not change the physics contents present on the model.

\section{Embedding the non-Abelian extension of the Proca model}      

The non-Abelian extension of the Proca model has its dynamics governed by the following Lagrangian density,

\be
\label{N0000}
{\cal L} = -\frac 14 F_{\mu\nu}^aF_a^{\mu\nu} + \frac 12 A_\mu^a A_a^\mu,
\ee
with

\be
\label{N0010}
F_{\mu\nu}^a = \partial_\mu A_\nu^a - \partial_\nu A_\mu^a + g C^a_{bc}A_\mu^bA_\nu^c,
\ee
where the antisymmetric tensor $C^a_{bc}$, $(C^a_{bc} = - C^a_{cb})$ are a set of real constants, known as the structure constants of the
gauge group, and satisfy a property, reads as

\be
\label{N0020}
C^a_{bc}C^d_{ae} + C^a_{eb}C^d_{ac} + C^a_{ce}C^d_{ab} = 0.
\ee

Since we are interested to analyze the non-Abelian Proca model from the symplectic point of view, the Lagrangian will be reduced to its
first-order form as follows,

\be
\label{N0030}
{\cal L}^{(0)} = \pi_a^i {\dot A}^a_i - \frac 12 (\pi_a^i)^2 + A_0^a\Omega_a - \frac 12 m^2 A_i^aA^i_a - \frac 12 m^2 A_0^aA^0_a -
\frac 14 F_{kj}^aF_{kj}^a,
\ee
where

\be
\label{N0040}
\Omega_a = \partial_i\pi^i_a - gC^b_{ca} \pi_b^iA_i^c + m^2A_a^0.
\ee
The symplectic variables and matrix are given by

\ba
\label{N0050}
\xi_\alpha^a &=& (A_i^a,\pi_i^a,A_0^a),\nonumber\\
f^{(0)} &=& \pmatrix{0 & - \delta_{ji}\delta^{ba} & 0\cr \delta_{ij}\delta^{ab} & 0  & 0\cr 0 & 0 & 0}\delta^{(3)}(\vec x - \vec y).
\ea
Since this matrix is singular, it has a zero-mode that generates the constraint $\Omega_a$, given in Eq.(\ref{N0040}). In agreement with
the symplectic method, this constraint is introduced into the kinetical sector of the first-order Lagrangian through a Lagrange multiplier,
namely,

\be
\label{N0060}
{\cal L}^{(1)} = \pi_a^i {\dot A}^a_i + \Omega_a\dot\eta^a - \frac 12 (\pi_a^i)^2 + A_0^a\Omega_a - \frac 12 m^2 A_i^aA^i_a -
\frac 12 m^2 A_0^aA^0_a - \frac 14 F_{kj}^aF_{kj}^a.
\ee
The extended symplectic variables are $\xi_\alpha^a = (A_i^a,\pi_i^a,A_0^a,\eta^a)$, and the symplectic matrix is

\be
\label{N0070}
f^{(1)} = \pmatrix{0 & - \delta_{ji}\delta^{ba} & 0 & - g C^{ab}_d\pi^d_i(y)\cr \delta_{ij}\delta^{ab} & 0  & 0 & \delta^{ab}\partial_i^y -
g C^{ab}_d A^d_i(y) \cr
 0 & 0 & 0 & m^2 \delta^{ab} \cr g C^{ba}_d\pi^d_j(x) & - \delta^{ba}\partial_j^x + g C^{ba}_d A^d_j(x) &
- m^2\delta^{ba} & 0}\delta^{(3)}(\vec x - \vec y).
\ee
This matrix is nonsingular and its inverse leads to the commutation relations among the dynamical variables, given by

\ba
\label{N0080}
\lbrace A_i^a(x), A_j^b(y)\rbrace &=& 0,\nonumber\\
\lbrace A_i^a(x), \pi_j^b(y)\rbrace &=& \delta^{ab}\delta_{ij}\delta^{(3)}(\vec x - \vec y),\nonumber\\
\lbrace \pi_i^a(x), \pi_j^b(y)\rbrace &=& 0,\\
\lbrace A_i^a(x), A_0^b(y)\rbrace &=& - \frac {1}{m^2} \delta^{ab}\partial_i^x \delta^{(3)}(\vec x - \vec y) -
\frac {g}{m^2} C^{ab}_e A^e_i(x)\delta(\vec x - \vec y),\nonumber\\
\lbrace A_0^a(x), A_0^b(y)\rbrace &=& - 2 gm^2C^{ab}_eA^e_0(x)\delta^{(3)}(\vec x - \vec y),\nonumber\\
\lbrace \pi_i^a(x), A_0^b(y)\rbrace &=& - \frac {g}{m^2} C^{ab}_e \pi^e_i(x)\delta^{(3)}(\vec x - \vec y).\nonumber
\ea
This completes the analysis of the noninvariant description of the model.

Afterhere, the model will be reformulated as a gauge invariant field theory. It will be done in the context of the symplectic
gauge-invariant formulation. In agreement with this formalism, the first-order Lagrangian (\ref{N0030}) will be rewritten as

\be
\label{N0090}
{\tilde {\cal L}}^{(0)} = \pi_a^i {\dot A}^a_i + \Psi_a\dot\theta^a - \frac 12 (\pi_a^i)^2 + A_0^a\Omega_a -
\frac 12 m^2 A_i^aA^i_a - \frac 12 m^2 A_0^aA^0_a - \frac 14 F_{kj}^aF_{kj}^a - G,
\ee
where the arbitrary functions are

\ba
\label{N0100}
\Psi_a &\equiv& \Psi_a(A_i^a,\pi_i^a,A_0^a,\theta^a),\nonumber\\
G &\equiv& G (A_i^a,\pi_i^a,A_0^a,\theta^a)=\sum_{n=0}^\infty{\cal G}^{n}(A_i^a,\pi_i^a,A_0^a,\theta^a),
\ea
where the $G$ function obeys a boundary condition, namely,

\be
\label{N0110}
G \equiv (A_i^a,\pi_i^a,A_0^a,\theta^a=0)={\cal G}^{0}(A_i^a,\pi_i^a,A_0^a,\theta^a=0)=0.
\ee
In this context, the corresponding symplectic matrix is

\be
\label{N0130}
f^{(0)} = \pmatrix{0 & - \delta_{ji}\delta^{ba} & 0 & \frac{\partial\Psi_b(y)}{\partial A^a_i(x)}\cr \delta_{ij}\delta^{ab} & 0  & 0 & \frac{\Psi_b(y)}{\partial \pi^a_i(x)} \cr 0 & 0 & 0 & \frac{\partial\Psi_b(y)}{\partial A^a_0(x)} \cr - \frac{\partial\Psi_a(x)}{\partial A^b_j(y)} & - \frac{\partial\Psi_a(x)}{\partial \pi^b_j(y)} & - \frac{\partial\Psi_a(x)}{\partial A^b_0(y)} & 0}\delta^{(3)}(\vec x - \vec y).
\ee

In order to determine the $\Psi_a$ function, we consider to analyze the symmetry related to the following zero-mode,

\be
\label{N0120}
\bar\nu^{(0)} = \pmatrix{\partial^x_i & 0 & 0 & 1}
\ee
with $\partial^x_i = \frac{\partial}{\partial x^i}$, which satisfies the following condition,

\be
\label{N0140}
\int_w\bar\nu^{(0)}(\vec x)f_{\alpha\beta}(\vec x - \vec w) = 0.
\ee
This condition produces a set of differential equations which allows us to compute the $\Psi_a$ function as

\be
\label{N0150}
\Psi_a =  - \partial_i \pi^i_a(x).
\ee
Consequently, the first-order Lagrangian is rewritten as

\be
\label{N0160}
{\tilde {\cal L}}^{(0)} = \pi_a^i {\dot A}^a_i - (\partial_i \pi^i_a)\dot\theta^a - {\tilde V}^{(0)},
\ee
where the symplectic potential is

\be
\label{N170}
{\tilde V}^{(0)} = \frac 12 (\pi_a^i)^2 - A_0^a\Omega_a + \frac 12 m^2 A_i^aA^i_a + \frac 12 m^2 A_0^aA^0_a + \frac 14 F_{kj}^aF_{kj}^a + G.
\ee
It completes the first step of the symplectic gauge-invariant formulation.

To unveil the hidden symmetry lying on the model, the zero-mode $\bar\nu^{(0)}$ does not generate a new constraint, consequently, we get a
relation, reads as

\be
\label{N0180}
\int_x \bar\nu^{(0)}_\alpha(w)\frac{\partial{\tilde V}(x)}{\partial \xi^a_\alpha(w)} = 0.
\ee
>From this relation we can compute the whole set of correction terms in order of $\theta^a$. Let us to start computing the linear
correction term in $\theta$, reads as

\be
\label{N0190}
\int_x \left\{\partial^w_i\frac{\partial V(x)^{(0)}}{\partial A_i^f(w)} + \frac{{\cal G}^{(1)}(x)}{\partial \theta^f(w)}\right\} = 0.
\ee
After an integration process, we get

\ba
\label{N0200}
{\cal G}^{(1)}(x) = &-& g C_{fa}^b \partial^x_i(A_0^a(x)\pi_b^i(x))\theta^f(x) - m^2 (\partial^x_iA^i_f)\theta^f(x)\nonumber\\
 &-& \frac 12 \int_y \partial^y_i \left(F^a_{kj}(x)\frac{\partial F_a^{kj}(x)}{\partial A_i^f(y)}\right)\theta^f(y).
\ea
Now, we will compute the quadratic term, namely,

\be
\label{N0210}
\int_x \left\{\partial^w_i\frac{\partial {\cal G}^{(1)}(x)}{\partial A_i^f(w)} + \frac{{\cal G}^{(2)}(x)}{\partial \theta^f(w)}\right\} = 0.
\ee
Integrating this relation in $\theta^f(w)$, the quadratic correction term is obtained as

\be
\label{N0220}
{\cal G}^{(2)}(x) =  \frac 12 m^2 (\partial_x^i \theta^f(x))^2 +
\frac 12\int_{\theta^f(x)}  \int_w \partial^w_i \int_y \left[(\partial^y_l{\cal A}^{il}_{fb})\theta^b(y)\right],
\ee
where

\be
\label{N0230}
{\cal A}^{il}_{fb} = \frac{\partial F^a_{kj}(x)}{\partial A_i^f(w)} \frac{\partial F_a^{kj}(x)}{\partial A_l^b(y)} +
F^a_{kj}(x)\frac{\partial^2 F_a^{kj}(x)}{\partial A_i^f(w)\partial A_l^b(y)}.
\ee

In view of this, two correction terms in order of $\theta_a$ $({\cal G}^{(3)}(x)$ and ${\cal G}^{(4)}(x))$ remain to compute. Let us to
start computing the first one. It can be done from the following relation,

\ba
\label{N0240}
\int_z\left\{\partial_n^z \left[\frac 12 \int_{\theta^f(x)}\int_w \partial_k^w \int_y \partial^y_l\frac{\partial{\cal A}^{kl}_{fb}}{\partial A_n^g(z)} \theta^b(y)\right] + \frac{{\cal G}^{(3)}(x)}{\partial \theta_g(z)} \right\} = 0\nonumber\\
{\cal G}^{(3)}(x) = - \frac 12 \int_{\theta^g(z)} \int_z \partial_n^z \int_{\theta_f(x)} \int_w \partial_k^w \int_y \partial^y_l\frac{\partial{\cal A}_{fb}^{kl}}{\partial A^n_g(z)} \theta^b(y).
\ea
To finish, the last correction term is computed as 

\be
\label{N0250}
{\cal G}^{(4)}(x) = \frac 12 \int_{\theta^h(v)} \int_v \partial_i^v \int_{\theta_g(z)} \int_z \partial_n^z \int_{\theta_f(x)} \int_w
\partial_k^w \int_y \partial^y_l\frac{\partial^2{\cal A}_{fb}^{kl}}{\partial A^i_h(v)\partial A^n_g(z)} \theta^b(y).
\ee
Therefore, the gauge invariant first-order Lagrangian is 

\be
\label{N0260}
{\tilde {\cal L}}^{(0)} = \pi_a^i {\dot A}^a_i - (\partial_i \pi^i_a)\dot\theta^a - {\tilde V}^{(0)},
\ee
where the gauge invariant Hamiltonian, identified as being the symplectic potential, is

\ba
\label{N270}
{\tilde {\cal H}} &=& \frac 12 (\pi_a^i)^2 - A_0^a\Omega_a + \frac 12 m^2 A_i^aA^i_a + \frac 12 m^2 A_0^aA^0_a + \frac 14 F_{kj}^aF_{kj}^a 
- g C_{fa}^b \partial^x_i(A_0^a(x)\pi_b^i(x))\theta^f(x)\nonumber\\ &-& m^2 (\partial^x_iA^i_f)\theta^f(x)
- \frac 12 \int_y \partial^y_i \left(F^a_{kj}(x)\frac{\partial F_a^{kj}(x)}{\partial A_i^f(y)}\right)\theta^f(y)
+ \frac 12 m^2 (\partial^i \theta^f(x))^2\nonumber\\
&+& \frac 12 \theta_f(w) \int_w \partial^w_i \int_y \left[(\partial^y_l{\cal A}_{fb}^{il})\theta^b(y)\right]
- \frac 12 \int_{\theta^g(z)} \int_z \partial_n^z \int_{\theta_f(x)} \int_w \partial_k^w \int_y \partial^y_l\frac{\partial{\cal A}_{fb}^{kl}}{\partial A^n_g(z)} \theta^b(y)\nonumber\\
&-& \frac 12 \int_{\theta^h(v)} \int_v \partial_i^v \int_{\theta_g(z)} \int_z \partial_n^z \int_{\theta_f(x)} \int_w
\partial_k^w \int_y \partial^y_l\frac{\partial^2{\cal A}_{fb}^{kl}}{\partial A^i_h(v)\partial A^n_g(z)} \theta^b(y).
\ea
This completes our proposal.

At this stage, we would like to reveal the hidden symmetry presents on the model from the Dirac's point of view. To this end, we start with the set of primary constraints, reads as

\ba
\label{N0280}
\Omega_1^a &=& \partial^i\pi_i^a + \pi_\theta^a,\nonumber\\
\chi_1^a &=& \pi_0^a.
\ea
For the first set of constraints, the temporal stability condition is satisfied $(\dot\Omega_1^a =0)$ , while for the second one, the following secondary constraints are required,

\be
\label{N0290}
\chi_2^a = \Omega^a - g C_f^{ba}\pi_b^i\partial_i\theta^f.
\ee
Due to this, the total Hamiltonian is 

\be
\label{N0300}
{\cal H} = {\tilde{\cal H}} + \lambda_a^1\Omega^a_1 + \zeta_a^1\chi_1^a + \zeta_a^2\chi_2^a,
\ee
where $\lambda_a^1$, $\zeta_a^1$ and $\zeta_a^2$ are Lagrange multipliers. Since the Poisson brackets among those constraints are

\ba
\label{N0310}
\lbrace\Omega^a_1(x),\Omega^b_1(y)\rbrace &=& 0,\nonumber\\
\lbrace\Omega^a_1(x),\chi^b_1(y)\rbrace &=& 0,\nonumber\\
\lbrace\Omega^a_1(x),\chi^b_2(y)\rbrace &=& 0,\\
\lbrace\chi^a_1(x),\chi^b_2(y)\rbrace &=& -m^2\delta^{ab}\delta^{(3)}(\vec x - \vec y),\nonumber\\
\lbrace\chi^a_2(x),\chi^b_2(y)\rbrace &=& 2gC_{d}^{ab} \chi_2^d(x)\delta^{(3)}(\vec x - \vec y) - 2 gm^2C_{d}^{ab}A_0^d(x)\delta^{(3)}(\vec x - \vec y),\nonumber
\ea
no more constraints arise. Note that some brackets above are null, indicating that there are both first and second class constraints.
Indeed, the first class constraint is $\Omega^a_1$ and second class ones are $\chi^a_1$ and $\chi^a_2$. In agreement with the Dirac's
procedure, the second class constraints can be taken equal to zero in a strong way, that allows us to compute the primary Dirac brackets.
Due to the Maskawa-Nakajima theorema\cite{NM}, the primary Dirac brackets among the phase space fields are canonical. To demonstrate, the
brackets are computed explicitly. The Dirac matrix is

\be
\label{N0320}
C = \pmatrix{0 & -m^2\delta^{cd} \cr m^2\delta^{dc} & B^{cd}}\delta^{(3)}(\vec x - \vec y),
\ee
with 

\be
\label{N0330}
B^{cd} = 2gC_{b}^{cd} \chi_2^b(x)- 2 gm^2C_{b}^{cd}A_0^b(x).
\ee
The inverse of Dirac matrix is

\be
\label{N0335}
C^{(-1)} = \frac {1}{m^2} \pmatrix{\frac {B^{cd}}{m^2} & \delta^{cd} \cr - \delta^{dc} & 0}\delta^{(3)}(\vec x - \vec y).
\ee
In accordance with the Dirac process, the Dirac brackets among the phase space fields are obtained as

\ba
\label{N0340}
\lbrace A^a_i(x), A^b_j(y)\rbrace^* &=& 0, \nonumber\\
\lbrace A^a_i(x), \pi^b_j(y)\rbrace^* &=& \delta^{ab}\delta^{(3)}(\vec x - \vec y), \nonumber\\
\lbrace A^a_i(x), A^b_0(y)\rbrace^* &=& -\frac {1}{m^2}\partial^x_i\delta^{(3)}(\vec x - \vec y) + \frac {g}{m^2} C^{ab}_f A_i^f(x)\delta^{(3)}(\vec x - \vec y), \nonumber\\
\lbrace \pi^a_i(x), A^b_0(y)\rbrace^* &=& -\frac{1}{m^2} g C^{ab}_e\pi^e_i\delta^{(3)}(\vec x - \vec y),\nonumber\\
\lbrace \pi^a_i(x), \pi^b_j(y)\rbrace^* &=& 0, \nonumber\\
\lbrace A^a_0(x), A^b_0(y)\rbrace^* &=& -\frac{g}{m^2} C_e^{ab}A^e_0(x)\delta^{(3)}(\vec x - \vec y),\\
\lbrace A^a_i(x), \theta^b(y)\rbrace^* &=& 0,\nonumber\\
\lbrace \pi^a_i(x), \theta^b(y)\rbrace^* &=& 0,\nonumber\\
\lbrace A^a_0(x), \theta^b(y)\rbrace^* &=& 0,\nonumber\\
\lbrace A^a_i(x), \pi_\theta^b(y)\rbrace^* &=& 0,\nonumber\\
\lbrace \pi^a_i(x), \pi_\theta^b(y)\rbrace^* &=& 0,\nonumber\\
\lbrace A^a_0(x), \pi_\theta^b(y)\rbrace^* &=& \frac {g}{m^2} C_e^{ab}\partial^x_i\pi^e_(x)\delta^{(3)}(\vec x - \vec y),\nonumber\\
\lbrace \theta^a(x), \pi_\theta^b(y)\rbrace^* &=& \delta^{ab}\delta^{(3)}(\vec x - \vec y),\nonumber\\
\lbrace \theta^a(x), \theta^b(y)\rbrace^* &=& 0,\nonumber\\
\lbrace \pi_\theta^a(x), \pi_\theta^b(y)\rbrace^* &=& 0.\nonumber
\ea
To accomplish the discussion, the infinitesimal gauge transformations are obtained, namely,

\ba
\label{N0350}
\delta A_i^a &=& - \partial_i^x\varepsilon^a,\nonumber\\
\delta \pi_i^a &=& 0,\nonumber\\
\delta A_0^a &=& 0,\\
\delta \theta^a &=& \varepsilon^a,\nonumber\\
\delta \pi_\theta^a &=& 0,\nonumber
\ea
which lead the Hamiltonian invariant.

To demonstrate that the gauge invariant formulation of the non-Abelian Proca model is dynamically equivalent to the original noninvariant
model, the symmetry is fixed by using the unitary gauge fixing procedure, reads as

\be
\label{N0360}
\varphi^a = \theta^a\approx 0,
\ee
which leads to the bracket below,

\be
\label{N0370}
\lbrace \Omega^a_1(x),\varphi^b(y)\rbrace = - \delta^{ab}\delta^{(3)}(\vec x - \vec y).
\ee
Due to this, a new Dirac brackets must be computed. The corresponding Dirac matrix for this set of constraints is

\be
\label{N0380}
C = \pmatrix{0 & -1 \cr 1 & 0 } \delta^{(3)}(\vec x - \vec y).
\ee
Using its inverse, the Dirac brackets among the physical phase space fields are computed, which is the same one calculated in the original
description given in Eq.(\ref{N0080}).

\section{Final Discussions}

In summary, we reformulate Abelian and non-Abelian noninvariant systems as gauge-invariant theories using a new gauge-invariant
formalism that is not affected by ambiguity problem related with the introduction of the WZ variables\cite{BN}. We start systematizing
the gauge-invariant formalism and, after, it was applied to a pedagogical model in order to illustrate and clarify some obscure points.
Afterward, we apply this formalism to an important Abelian models, NLSM and CSM. In the former, a hidden symmetry lying on the original
phase space was disclosed, oppositely to other approaches\cite{WN,BGB,BN}, where the symmetry resides on the extended WZ phase space. In
the later, the chiral anomaly was canceled and the gauge symmetry was restored. It is important to notice that it was achieved introducing
one WZ field while other schemes have success with the introduction of two or more WZ fields, which is the origin of the ambiguity problem.
Further, we showed, in the context of a simple non-Abelian model (the non-Abelian Proca model) that the symplectic gauge-invariant
formalism can be used without any restrictions with the algebra obeyed by the noninvariant model, while other constraint conversion
techniques work since the algebra is previously, and necessarily, taken in account. 

\section{ Acknowledgements}
This work is supported in part by FAPEMIG and CNPq, Brazilian Research Agencies.  In particular, C. Neves would like to  acknowledge the
FAPEMIG grant no. CEX-00005/00 and A.C.R.Mendes would like to thanks to CNPq for partial support.

\end{document}